==
\documentclass[aps,pra,onecolumn,notitlepage,10pt]{revtex4-1}

\usepackage{braket}
\usepackage{amsmath}
\usepackage{color}
\usepackage{graphicx}
\usepackage{subcaption}
\usepackage{caption}
\usepackage{amssymb}
\usepackage{amsfonts}
\usepackage{natbib}

\begin{document}


\title{Systematic Dimensionality Reduction for Quantum Walks: \\Optimal Spatial Search and Transport on Non-Regular Graphs}


\author{Leonardo Novo$^{1,2,a}$}  
\email[Correspondence to: ]{lnovo@lx.it.pt}
\author{Shantanav Chakraborty$^{1,2,a}$}
\author{Masoud Mohseni$^3$} 
\author{Hartmut Neven$^3$}
\author{Yasser Omar$^{1,2,4}$}
\affiliation{$^1$Physics of Information Group, Instituto de Telecomunica\c{c}\~oes, Portugal}
\affiliation{$^2$Instituto Superior T\'{e}cnico, Universidade de Lisboa, Portugal}
\affiliation{$^3$Google Inc., Venice, CA 90291, USA}
\affiliation{$^4$CEMAPRE, ISEG, Universidade de Lisboa, Portugal}
\collaboration{$^a$ Both authors have equal contribution}
\date{2 November 2015}

\begin{abstract}
Continuous time quantum walks provide an important framework for designing new algorithms and modelling quantum transport and state transfer problems. Often, the graph representing the structure of a problem contains certain symmetries that confine the dynamics to a smaller subspace of the full Hilbert space. In this work, we use invariant subspace methods, that can be computed systematically using Lanczos algorithm, to obtain the reduced set of states that encompass the dynamics of the problem at hand without the specific knowledge of underlying symmetries. First, we apply this method to obtain new instances of graphs where the spatial quantum search algorithm is optimal: complete graphs with broken links and complete bipartite graphs, in particular, the star graph. These examples show that regularity and high-connectivity are not needed to achieve optimal spatial search. We also show that this method considerably simplifies the calculation of quantum transport efficiencies. Furthermore, we observe improved efficiencies by removing a few links from highly symmetric graphs. Finally, we show that this reduction method also allows us to obtain an upper bound for the fidelity of a single qubit transfer on an XY spin network.      

\end{abstract}
\pacs{}
\maketitle
\section*{Introduction}
Quantum walks \cite{zagury, farhidectree, kempeqw, aharonov, kendon, reitznerreview, venegas} are an important framework to model quantum dynamics, with applications ranging from quantum computation to quantum transport. Being the quantum mechanical analogue of classical random walks, quantum walks can outperform their classical counterparts by exploiting interference in the superposition of the various paths in a graph as well as by taking advantage of quantum correlations and quantum particle statistics between multiple walkers \cite{yasser2w}. In fact, for multiparticle quantum walks, interactions lead to an efficient simulation of the circuit model of quantum computation \cite{childsmultiversal}.

Quantum walks can be formulated in both the discrete time \cite{zagury} and continuous time \cite{farhidectree} frameworks, where the latter can be obtained as a limit of the former \cite{childsDTvsCT}. In this article we focus on single particle continuous time quantum walks (CTQWs), where the knowledge of the adjacency matrix of the graph is sufficient to completely describe the walk. Several interesting algorithms have been developed in this framework \cite{Childs_grover, childsglued}. In fact, CTQWs in sparse unweighted graphs are equivalent to the circuit model of quantum computation, although the corresponding simulation is not efficient \cite{childs_universal}. Besides quantum algorithms, CTQWs are applied in areas such as quantum transport \cite{mohseniqw, masoud_enaqt, pleniodat, caruso_enaqt} and state transfer \cite{sougato_stransfer,kay}. 

In most CTQW problems, the quantity of interest is the population (or probability amplitude) at a particular node of the graph. For example, in the spatial search algorithm \cite{Childs_grover}, the purpose is to maximize the probability amplitude at the solution node in the shortest possible time. In the glued trees algorithm demonstrated in \cite{childsglued}, which has an exponential speed up over its classical counterpart, the walker traverses the graph therein in order to find an exit node. On the other hand, quantum transport problems involving a single excitation, for example, exciton transport in photosynthetic complexes, can be modelled by CTQWs \cite{mohseniqw, masoud_enaqt, pleniodat, caruso_enaqt, qbiobook}. In such cases, the figures of merits are typically the transport efficiency or the transfer time to a special node, known as the \textit{trap}. As for state transfer, the task is to send a qubit from one point of a spin-network to another with maximum fidelity \cite{sougato_stransfer, kay, severini}. In many of these problems, the graph in which the walk takes place possesses some symmetry \cite{Meyer_symmetry, Childs_grover, Farhi_analog_grover} which implies that the dynamics of the walker is restricted to a subspace that is smaller than the complete Hilbert space spanned by the nodes of the graph. In this work, we use invariant subspace methods, that can be computed systematically using Lanczos algorithm \cite{lanczos}, to obtain a reduced model that fully describes the evolution of the probability amplitude at the node we are interested in. This method involves obtaining the minimal subspace which contains this node and is invariant under the unitary evolution. This is simply the subspace that contains the node of interest $\ket{w}$, and all powers of the Hamiltonian applied to it, also known as a Krylov subspace. Henceforth, this subspace will be denoted by $\mathcal{I}(H,\ket{w})$. This subspace can be systematically obtained without taking into consideration the symmetries of the Hamiltonian, using, for example, Lanczos algorithm \cite{lanczos}. This algorithm iteratively obtains the basis for the invariant subspace: the first basis element is the special node $\ket{w}$; the $i$th basis element is calculated by applying the Hamiltonian to the $(i-1)$th element and orthonormalizing with respect to the previous basis elements. When expressed in the Lanczos basis, any Hermitian matrix becomes a tridiagonal matrix. Thus, \textit{any problem in quantum mechanics} wherein the dynamics is described by a time independent Hamiltonian can be mapped to a CTQW on a weighted line, where the nodes are the elements of the Lanczos basis. In this way, we explore the notion of invariant subspaces to systematically reduce the dimension of the Hamiltonian that completely describes the  dynamics relevant to our problem. We use this method to obtain new results on several CTQW problems, as well as re-derive some other known results in a simpler manner.
 
First, we consider the spatial search algorithm \cite{Childs_grover}, which searches for an element contained in one of the $N$ nodes of the graph in $\mathcal{O}(\sqrt{N})$ time, which is optimal \cite{Farhi_analog_grover}. This algorithm is known to hold optimally for structures such as the complete graph, hypercubes, lattices of dimension greater than four and more recently, for strongly regular graphs \cite{Meyer_symmetry}. In two dimensional lattices, the lower bound of $\mathcal{O}(\sqrt{N}\text{log} N)$ could only be achieved when the dispersion relation of the spectrum is linear at a certain energy, i.e., it contains a Dirac point, as in honeycomb (e.g. graphene) lattices \cite{tanner_graphene}, and   crystal lattices \cite{childs_2d_search}. However, the characteristics that a graph must possess, in general, for this algorithm to run optimally remains an open question. In fact in \cite{aaronson}, where the authors present a different spatial search algorithm based on the divide and conquer approach, their main criticism towards the CTQW version of the spatial search was the fact that an upper bound on the running time is unknown even if ``minor defects are introduced''. Here, we show that the algorithm runs optimally on the complete graph with imperfections in the form of broken links, and also for complete bipartite graphs (CBG). In both cases, the graphs are, in general non-regular, i.e. not all the nodes of the graph have the same degree. A particular case of the CBG is the star graph where $N-1$ nodes are connected only to a central node, which is a planar structure with link connectivity one. Thus, this example shows that high connectivity is not a requirement for optimal quantum search. Moreover, on removing $k$ links, such that $k<<N$, from a star graph, the emerging graph is robust as it preserves its star connectivity and search is still optimal provided that the broken link does not contain the solution. In all the graphs mentioned thus far, the Hamiltonian of dimension $N$,  describing the dynamics of the algorithm, can be reduced to a Hamiltonian of dimension at most four. The dynamics, driven by this reduced Hamiltonian, can be viewed as a CTQW on a smaller graph, which provides an intuitive picture of the algorithm, similar to a quantum transport problem. It is worth noting that the reduced Hamiltonians presented here describe the dynamics of the problem exactly and are not obtained by approximating the search Hamiltonian at the avoided crossing as in \cite{tanner_graphene}. Thus, this is a simple way to analyse the algorithm that, in some cases, allows us to understand why search is optimal in a certain graph without having to explicitly calculate the eigenstates of the Hamiltonian.

Furthermore, we consider quantum transport on a graph, where an exciton is to be transferred from one node to a special node where it gets absorbed, known as the trap \cite{masoud_enaqt, caruso_enaqt}. In the scenario where there is no disorder, decoherence or losses, it was shown in \cite{caruso_enaqt} that the transport efficiency is given by the overlap of the initial condition with the eigenstates having a non-zero overlap with the trap. We prove that this subspace is the same as the invariant subspace $\mathcal{I}(H,\ket{trap})$. This observation allows us to compute transport efficiencies without having to diagonalize the Hamiltonian. We calculate the efficiency in the complete graph (CG) with this method (obtaining the same result as in \cite{caruso_enaqt}, which uses the eigenstates of the graph). Furthermore, we obtain the transport efficiency on binary trees and hypercubes as a function of the number of generations and dimension respectively, for various initial conditions. Finally, we show that the efficiency in all these structures increases on average, when a few links are broken randomly from the graph. A particularly interesting example is the one of breaking the link from the complete graph which connects the initial and trap nodes. In such a case, the efficiency increases to 1, in the absence of decoherence and losses, irrespective of the size of the network. For this case, we also calculate analytically the trapping time, which does not depend on $N$. This counter-intuitive result can be interpreted by looking at the reduced subspace of the graph, where the problem reduces to an end to end transport in  a line of three nodes. Similar results were obtained in \cite{severini}, in the context of state transfer, although different methods were used for the analysis of the problem. Overall, the instances presented herein show that even small perturbation to the symmetry of a structure leads to a drastic improvement of the transport efficiency in the absence of decoherence. When decoherence is present, the effect of geometry in the transport efficiency was numerically studied in \cite{geometrical}, for random disordered structures.
 
Finally, we connect the results obtained for transport to the problem of state transfer in a quantum network. In the single excitation framework, the state transfer problem is equivalent to a CTQW of a single particle. We show that the fidelity of transferring an excitation from one node of the network $i$ to another node $j$, is upper bounded by the square root of the transport efficiency in the analogous transport problem wherein $i$ is the initial state and $j$ is the trap node. This gives a simple way to upper bound the fidelity of transferring a qubit in any spin network.

Overall, we demonstrate that dimensionality reduction using the notion of invariant subspaces can be a useful tool to analyse CTQW based problems. By mapping a QW problem on a graph to one on a much smaller structure, the analysis of the problem becomes easier and the dynamics of the walk can be intuitively understood. Krylov subspace methods and the Lanczos algorithm for the analysis of CTQWs have also been used in \cite{zafar}, but different results were obtained therein.  In the discrete time framework, the role of symmetry and invariant subspaces were studied in \cite{brun, daniel}. Krylov subspace approaches were used to analyse adiabatic quantum search on structured database in \cite{sufiani}, and to obtain bounds for information propagation on lattices in \cite{gong}. The notion of invariant subspaces is also exploited in \cite{mohan} to simplify the analysis of parametrized Hamiltonians of quantum many body systems.

This paper is structured as follows: In Sec.~\textit{Methods}, the systematic method to obtain the reduced subspace $\mathcal{I}(H,\ket{\omega})$ is demonstrated. Sec.~\textit{Results} comprises of the various applications of the reduction method, namely in quantum spatial search, quantum transport and state transfer. Finally, we present our conclusions in Sec. \textit{Discussions}.  
\section*{Methods}\label{red_Ham}
\textbf{\textit{Dimensionality reduction of continuous time quantum walks}:} Let us consider a graph $G(V,E)$ of $N$ nodes, where $V$ is the set of nodes and $E$, the set of links. The adjacency matrix $A$ of $G(V,E)$ is of dimension $N\times N$ and is defined as follows: 
\begin{equation}
 A_{ij} = \left\{ 
  \begin{array}{l l}
    1 & \quad \text{if $(i,j)\in $ E}\\
    0 & \quad \text{otherwise}
  \end{array} \right.~ .
  \end{equation}
Formally, a CTQW on the graph $G(V,E)$ takes place on a Hilbert space $\mathcal{H}$ of dimension $N$ that is spanned by the nodes of the graph $\ket{i}$ with $i\in V$. A particle starting in a state $|\psi_0\rangle\in \mathcal{H}$ evolves according to the Schr\"{o}dinger equation where the Hamiltonian $H$ that governs the system dynamics is the adjacency matrix, i.e., $H=A$. After time $t$, the particle is in the state
\begin{equation}
\ket{\psi(t)}=e^{-iHt}\ket{\psi_0},
\end{equation}
and the probability that the walker is in node $v$ is given by $|\braket{v|e^{-iHt}|\psi_0}|^2$. The unitary evolution of the state $\ket{\psi_0}$ can be expressed as 
\begin{equation}\label{eq:unitary}
\ket{\psi(t)}= U(t)\ket{\psi_0} = e^{-iHt}\ket{\psi_0} = \sum_{k=0}^{\infty}\dfrac{(-i t)^k}{k!}H^k\ket{\psi_0}.
\end{equation}
So $\ket{\psi(t)}$ is contained in the subspace $\mathcal{I}(H,\ket{\psi_0})=\text{span}(H^k \ket{\psi_0}),~\forall k\in \mathbb{N}$. This subspace of $\mathcal{H}$ is invariant under the action of the Hamiltonian and thus also of the unitary evolution. Trivially, the dimension of this subspace is at most $N$. However, if the Hamiltonian is highly symmetrical, only a small number of powers of $H^k \ket{w}$ are linearly independent and the dimension of $\mathcal{I}(H,\ket{\psi_0})$ can be much smaller than $N$. Thus, we can reduce the dimension of the problem in the following way. Let $P$ be the projector onto $\mathcal{I}(H,\ket{\psi_0})$. Then
\begin{align}
U(t)\ket{\psi_0}&=P U(t) P \ket{\psi_0}=\\ \nonumber
&= \sum_{k=0}^{\infty}\dfrac{(-i t)^k}{k!}(P H P)^k \ket{\psi_0}\\ \nonumber
&= e^{-i P H P t} \ket{\psi_0}=e^{-i H_{\text{red}} t} \ket{\psi_0},
\end{align} 
where $H_{red}=P H P$ is the reduced Hamiltonian. In the derivation we used the fact that $P^2=P$, $P \ket{\psi_0}=\ket{\psi_0}$ and $P U(t) \ket{\psi_0}=U(t) \ket{\psi_0}$. Now, for any state $\ket{\phi}\in\mathcal{H}$, we have
\begin{align}
\bra{\phi}U(t)\ket{\psi_0}&=\bra{\phi} P P U(t) P \ket{\psi_0}\\ \nonumber
&=\bra{\phi}P e^{-i H_{\text{red}} t} \ket{\psi_0}\\
&= \bra{{\phi}_{\text{red}}} e^{-i H_{\text{red}} t} \ket{\psi_0} ,
\end{align} 
where, the reduced state, $\ket{{\phi}_{\text{red}}}= P \ket{{\phi}}$. The same reasoning could be applied using the projector $P'$ onto the subspace $\mathcal{I}(H,\ket{\phi})$, in which case we obtain  
\begin{align}
\bra{\phi}U(t)\ket{\psi_0}= \bra{{\phi}} e^{-i H'_{\text{red}} t} \ket{{\psi_0}_{\text{red}}},
\end{align}
with $H'_{red}=P' H P'$ and $\ket{{\psi_0}_{\text{red}}}= P' \ket{{\psi_0}}$.
This way, a reduced Hamiltonian can be obtained which can be seen as the Hamiltonian of a weighted graph that in some cases, can be much simpler than the original graph we started with. Here, $\ket{\phi}$ can be the solution node for search algorithms, the trap for quantum transport and the target node for state transfer problems (see Sec.~\textit{Results}).  

A systematic way to calculate an orthonormal basis of $\mathcal{I}(H,\ket{\phi})$ is given by Lanczos algorithm \cite{lanczos}. This basis, which we denote by $\{\ket{l_1},\dots,\ket{l_m}\}$, can be obtained as follows:
the first element is $\ket{l_1}=\ket{\phi}$; the $k^{\text{th}}$ element $\ket{l_k}$ is obtained by orthonormalizing $H^k\ket{\phi}$ with respect to the subspace spanned by $\{\ket{l_1},\dots, \ket{l_{k-1}}\}$. The procedure stops when we find the minimum $m$ such that $H^{m+1}\ket{\phi}\in\text{span}(\{\ket{l_1},\dots,\ket{l_m}\})$. It can be shown that $H$ projected in this basis has a tridiagonal form:
\begin{equation}
H_{red}= \left(
\begin{array}{c c c c c}
E_1&V_2&0&\dots&0\\
V_{2} & E_2& V_3&\dots&0\\
0   & V_3&E_3&\ddots&\vdots\\
\vdots&~& \ddots& \ddots&V_m\\
0&\dots&0&V_{m}& E_m
\end{array}
\right).
\end{equation}
This implies that, in fact, any problem in quantum mechanics with a time independent Hamiltonian can be mapped to an equivalent problem governed by a tight-binding Hamiltonian of a line with $m$ sites, with site energies $E_i$ and couplings $V_i$.
 
\begin{figure}
\begin{subfigure}{0.5\textwidth}
\includegraphics[scale=0.6]{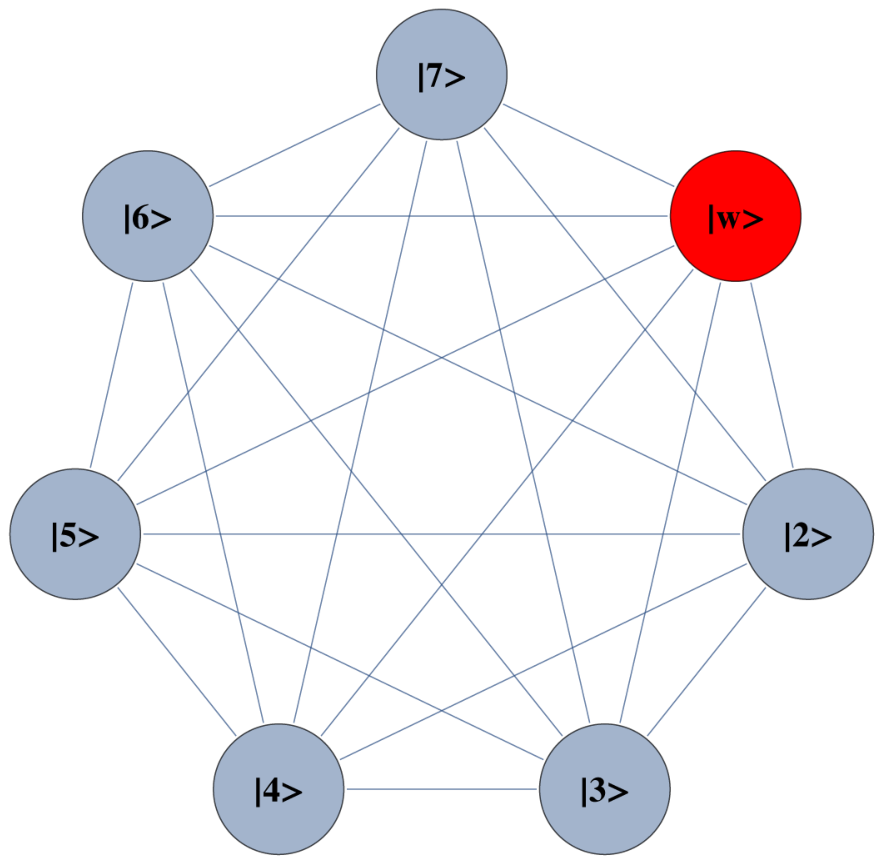}
\caption{}
\label{complete}
\end{subfigure}
\\~
\vspace{0.5cm}
\\
\begin{subfigure}{0.5\textwidth}
\includegraphics[scale=0.8]{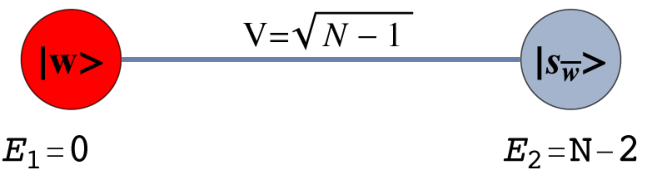}
\caption{}
\label{red_completegraph}
\end{subfigure}
\\~
\vspace{0.5cm}
\\
\begin{subfigure}{0.5\textwidth}
\includegraphics[scale=0.8]{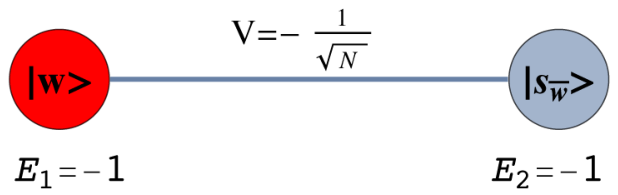}
\caption{}
\label{red_completegraph_search}
\end{subfigure}
\caption{
a) A complete graph with 7 nodes, with the target node in red. b) Line with two nodes representing the reduced Hamiltonian of the complete graph with N nodes (see Eq.\eqref{redH_completegraph}), The solution node $\ket{w}$ is represented in red and the other node $\ket{s_{\bar{w}}}$ represents the equal superposition of all nodes except $\ket{w}$. $E_1$ and $E_2$ represent their respective site energies and V the coupling between them. c) The search Hamiltonian in the reduced picture (see Eq.\eqref{Ham_line2d}). In contrast to Fig.~1b, the site energies of both the nodes are equal, leading to perfect transport between them. Also, the transport time is given by the inverse of the coupling, which yields the running time of the algorithm $T=\mathcal{O}(\sqrt{N})$.
}

\end{figure}

To illustrate the reduction method, we give a simple example of the quantum walk on the complete graph of N nodes, a graph wherein every node is connected to every other node, as shown in Fig.~\ref{complete}. 
The Hamiltonian is given by the adjacency matrix of the graph 
\begin{equation}
H= \sum_{i\neq j} \ket{i}\bra{j}.
\end{equation}
In this case, if $\ket{w}$ is a node of the graph, we have that
\begin{align}
H\ket{w}=\sqrt{N-1}\sum_{i\neq w}\dfrac{1}{\sqrt{N-1}}\ket{i}=\sqrt{N-1}\ket{s_{\bar{w}}},
\end{align}
where we define the equal superposition of all nodes except $\ket{w}$ as 
\begin{equation}
\label{swbar}
\ket{s_{\bar{w}}}=\frac{1}{\sqrt{N-1}}\sum_{i\neq w}\ket{i}.
\end{equation}
Furthermore, 
\begin{equation}
H\ket{s_{\bar{w}}}= (N-2)\ket{s_{\bar{w}}}+\sqrt{N-1}\ket{w}.
\end{equation}
It is easy to see now that any state $H^k\ket{w}$ can be written as a linear combination of $\ket{w}$ and $\ket{s_{\bar{w}}}$. Thus, to calculate $U(t)\ket{w}$, it is enough to consider the dynamics in this two dimensional subspace spanned by $\{\ket{w},\ket{s_{\bar{w}}}\}$ driven by the reduced Hamiltonian
\begin{equation}\label{redH_completegraph}
H_{red}=\left(\begin{array}{c c}
0 & \sqrt{N-1} \\
\sqrt{N-1}& N-2\end{array}\right).
\end{equation}   
This approach reduces the problem of calculating $U(t)\ket{w}$ to the calculation of the exponential of a $2\times2$ matrix instead of a $N\times N$ matrix. We can see $H_{red}$ as a tight-binding Hamiltonian of a structure with two sites $\ket{w}$ and $\ket{s_{\bar{w}}}$, with site energies $0$ and $N-2$, respectively and a coupling of $\sqrt{N-1}$, as shown in Fig. \ref{red_completegraph}.   
Another interesting example is the reduction of the quantum walk on the glued trees of height $l$ with $\mathcal{O}(2^l)$ nodes to the $2l+1$ column states as done in \cite{childsglued}, which is crucial to prove the exponential speed-up of this algorithm with respect to its classical counterpart. Even when some symmetry of the graph is broken, say by breaking a link of the graph, this reduction is still very useful and captures the symmetry that remains (see Sec.~\textit{Results}).

This reduction method can also be used in the context of quantum transport. In fact, in the \textit{Supplementary Information}, we show that $\mathcal{I}(H,\ket{w})$ is equal to the subspace spanned by the eigenstates of the Hamiltonian which have a non-zero overlap with $\ket{w}$. This subspace is referred to as the `non-invariant subspace' in \cite{caruso_enaqt} where $\ket{w}$ is the trapping site. Let us denote this subspace as $\Lambda(H,\ket{w})$. The calculation of $\Lambda(H,\ket{w})$ is important for computing the transport efficiency in various networks, in the absence of interaction with the environment. The Lanczos' method provides a simpler way to calculate this subspace which eliminates the need to compute the eigenstates of the full Hamiltonian. This way, it also enables us to efficiently analyse the effects of perturbing the symmetry of networks in transport dynamics, as described in Subsec.~\emph{Applications to Quantum Transport} of \textit{Results}. 

In the following section, we use this method to analyse spatial search in highly symmetric graphs, calculate efficiency of transport in several structures and obtain bounds on the fidelity of single qubit state transfer in spin networks.
\section*{Results}\label{Grover}
\subsection*{Applications to spatial search}
The goal of the spatial search algorithm in the CTQW formalism is to find a marked basis state $|w\rangle$ \cite{Childs_grover, Meyer_symmetry} and proceeds by evolution of the initial state $|s\rangle=\frac{1}{\sqrt{N}}\sum_{i} |i\rangle$, according to the Hamiltonian
\begin{equation}
H=-\gamma A-|w\rangle\langle w|,
\label{Grover_Ham}
\end{equation}
where $A$ is the adjacency matrix of $G$ and $\gamma$ is the coupling between connected nodes that is tuned so as to run the algorithm optimally.
As described in Sec.~\textit{Methods}, the Hamiltonian of a complete graph can be reduced to a two dimensional subspace, which can be seen as a line with two nodes. The reduced Hamiltonian is given by
\begin{equation}\label{Ham_line2d}
H_{\text{search}}=-\gamma
\begin{bmatrix}
\frac{1}{\gamma} & \sqrt{N-1}   \\ 
\sqrt{N-1} & N-2 
\end{bmatrix}
\end{equation}
and is depicted in Fig.~\ref{red_completegraph_search}. The optimal value of $\gamma$ is proven to be $1/N$ such that the dynamics is simply a rotation between $\ket{s}$ and $\ket{w}$ \cite{Childs_grover}. This value is optimal because it ensures that the site energies of both the nodes $\ket{w}$ and $\ket{s_{\bar{w}}}$ are equal, thus optimizing transport between these nodes. The initial state is approximately $\ket{s_{\bar{w}}}$ so the probability amplitude at $\ket{w}$ becomes approximately 1 after a time that is of the order of the inverse of the coupling. Hence, the running time of the algorithm is $\mathcal{O}(\sqrt{N})$. 
Here, we give examples of non-regular graphs where the algorithm runs optimally, by making use of the reduction method explained in Sec.~\textit{Methods}. First, we analyse the effect of breaking links from this graph and show that the optimal running time is maintained. This can be interpreted as an inherent robustness of the algorithm to imperfections of this form. Furthermore, we prove that the spatial search algorithm also runs optimally for complete bipartite graphs.

\vspace{.5cm} 
\textbf{\textit{Optimal spatial search on complete graphs with broken links}: }
\begin{figure}[h]
\centering
\begin{subfigure}{0.3\textwidth}
\centering
\includegraphics[scale=0.8]{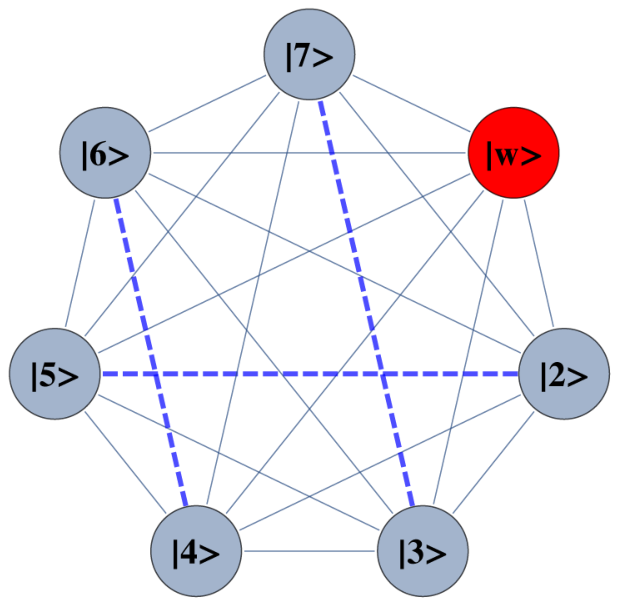}
\caption{}
\label{completegraph_kBL}
\end{subfigure}
\\~
\vspace{0.5cm}
\\
\begin{subfigure}{0.6\textwidth}
\centering
\includegraphics[scale=0.05]{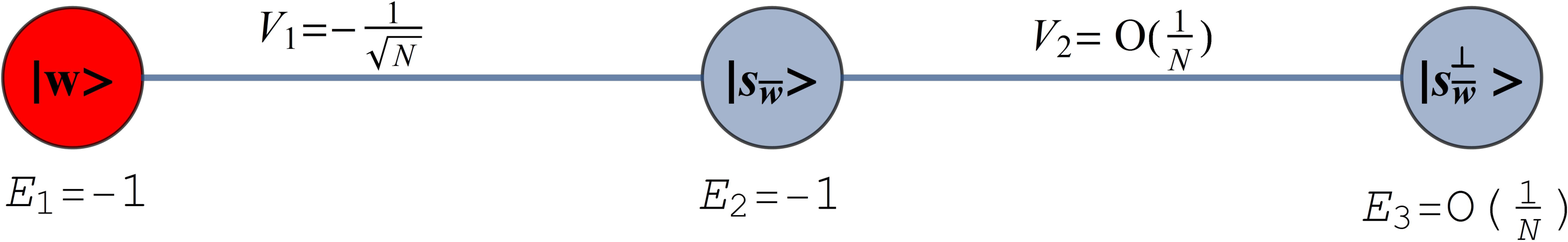}
\caption{}
\label{red_completegraph_kBL}
\end{subfigure}
\caption{ a) Complete graph with 7 nodes and 3 broken links (dashed links). No more than one link per node is broken. b) Representation of the reduced search Hamiltonian for the complete graph with $N$ nodes and $k$ broken links where at most one link is broken per node. The coupling $V_2$ to the third node $\ket{s_{\bar{w}}^{\perp}}$ is much weaker than the coupling $V_1$ and can be neglected. Thus, the dynamics is the same as in Fig.~1c.}
\end{figure} 
Here, we consider the case of breaking $k$ links from a complete graph and show analytically that the optimality of the algorithm is maintained. We assume that at most one link is removed per node and hence 
\[ k \leq \left\{ 
  \begin{array}{l l}
    N/2 & \quad \text{, if $N$ is even}\\
    (N-1)/2 & \quad \text{, if $N$ is odd.}
  \end{array} \right.\]
In this scenario, there exist two cases that require separate analytical treatment, namely one where none of the broken $k$ links were connected to the solution state $\ket{w}$, and the other where one of the broken link was connected to $\ket{w}$. We analyse the former in this section while the latter is explained in the \textit{Supplementary Information}.    

Let us consider that the links broken correspond to the set $E_{\text{broken}}=\{(i_1,~i_2), \cdots, (i_{2k-1},~i_{2k})\}$, that is, at most one link is removed from each node, as shown in Fig.~\ref{completegraph_kBL}. Also, let $V_{\text{broken}}=\{i_1,~i_2, \cdots, i_{2k-1},~i_{2k}\}$ be the set of nodes comprising of the broken links. The graphs obtained by breaking links from a complete graph are not regular and hence violates the requirement for regularity in networks in order to achieve a quadratic speed up. 
Applying Lanczos algorithm, we obtain the reduced basis $\mathcal{I}(H,\ket{w})=\text{span}(\ket{w}, \ket{s_{\bar{w}}}, \ket{s_{\bar{w}}^\perp})$, where $\ket{s_{\bar{w}}}$ is defined as 
\begin{equation}
\ket{s_{\bar{w}}}=\frac{1}{\sqrt{N-1}}\sum_{q \neq w}{\ket{q}},
\end{equation}
i.e., the equal superposition of all nodes of the graph except the solution and thus $\ket{s_{\bar{w}}}\approx\ket{s}$. $\ket{s_{\bar{w}}^\perp}$ is a state that is orthogonal to both $\ket{w}$ and $\ket{s_{\bar{w}}}$ and is constructed as,
\begin{equation}\label{sup_rest}
\ket{s_{\bar{w}}^\perp}=\sqrt{\frac{2k}{N-1}}\ket{s_{\bar{k}}}-\sqrt{\frac{N-2k-1}{N-1}}\ket{s_k},
\end{equation}
where $\ket{s_k}=(1/\sqrt{2k})\sum_{j\in V_{\text{broken}}}\ket{j}$ and $\ket{s_{\bar{k}}}=(1/\sqrt{N-2k-1})\sum_{j\notin (V_{\text{broken}},w)}\ket{j}$.

Also, let $k=\alpha N$ such that $\alpha \in \mathbb{R}$, and, $0\leq \alpha \leq \frac{1}{2}-\frac{1}{N}$. Thus, for large $N$, the search Hamiltonian in this basis is,
\begin{equation}
H_{\text{search}}= -\gamma
\begin{bmatrix}
\frac{1}{\gamma} & \sqrt{N} & 0 \\
\sqrt{N} & N & \sqrt{2\alpha(1-2\alpha)}\\
0 & \sqrt{2\alpha(1-2\alpha)} & 2\alpha - 2 
\end{bmatrix}.    
\end{equation}
We shall use degenerate perturbation theory to estimate the running time of the algorithm in this scenario \cite{Meyer_symmetry}. We write, $H_{\text{search}}=H^{(0)}+H^{(1)}+H^{(2)}$ with
\begin{align}
H^{(0)}=& 
\begin{bmatrix}
-1 & 0 & 0   \\ 
0 & -\gamma N & 0 \\
0 & 0 & 0 
\end{bmatrix},\\
H^{(1)}=&\begin{bmatrix}
0 &-\gamma \sqrt{N} & 0   \\ 
-\gamma\sqrt{N} & 0 &0 \\
0 &0 & 0 
\end{bmatrix},\\
H^{(2)}=&
\begin{bmatrix}
0 & 0 & 0 \\
0 & 0 & -\gamma\sqrt{2\alpha(1-2\alpha)} \\
0 &-\gamma\sqrt{2\alpha(1-2\alpha)}  & -\gamma (2\alpha -2)
\end{bmatrix},
\end{align}
such that $H^{(0)}$ has terms of $\mathcal{O}(1)$, $H^{(1)}$ has terms of order $\mathcal{O}(\frac{1}{\sqrt{N}})$ while $H^{(2)}$ contains $\mathcal{O}(\frac{1}{N})$ terms (see Fig.\ref{red_completegraph_kBL}). $H^{(0)}$ has eigenstates $\ket{w}$ and $\ket{s_{\bar{w}}}$ with eigenvalues $-1$ and $-\gamma N$ respectively and thus, in order for the dynamics to rotate between $|s\rangle$ (which is approximately $\ket{s_{\bar{w}}}$ for large $N$) and the solution state $\ket{w}$, the eigenvalues must be degenerate, making $\gamma=\frac{1}{N}$. The eigenstates of the perturbed Hamiltonian are $\ket{\lambda_\pm}=\frac{1}{\sqrt{2}}(\ket{s_{\bar{w}}}\mp\ket{w})$ having eigenvalues $E_\pm=-1\pm\frac{1}{\sqrt{N}}$. This gives the running time of the algorithm to be
\begin{equation}
T=\frac{\pi}{\Delta E}=\frac{\pi\sqrt{N}}{2},
\end{equation}
thereby preserving the optimal quadratic speed up.

\begin{figure}[h]
\centering
\begin{subfigure}{0.5\textwidth}
\centering
\includegraphics[scale=0.7]{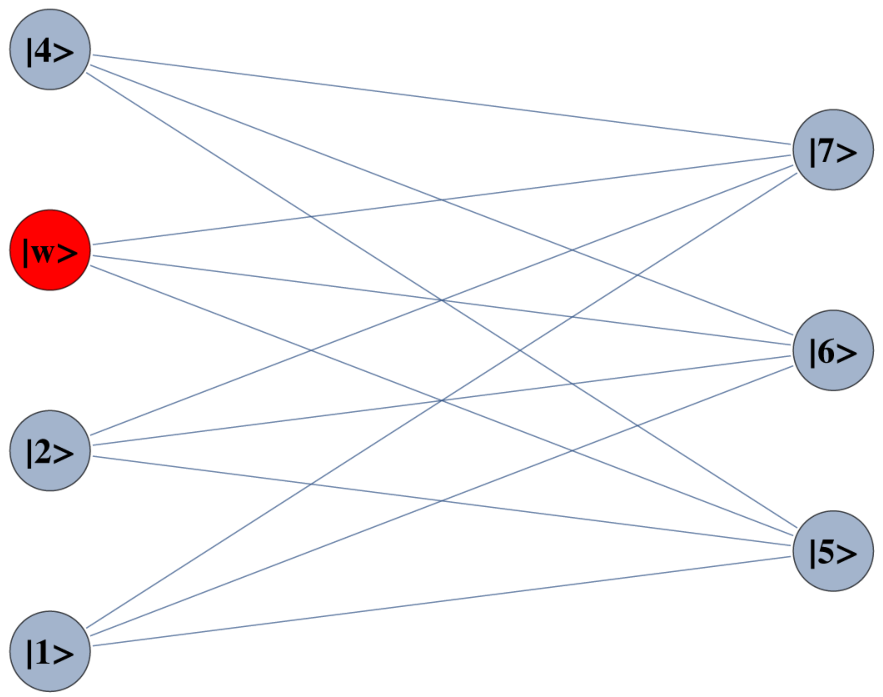}
\caption{}
\label{completebigraph}
\end{subfigure}
\\~
\vspace{0.5cm}
\\
\begin{subfigure}{0.5\textwidth}
\centering
\includegraphics[scale=0.65,trim= 0cm 0pt 0pt 0pt]{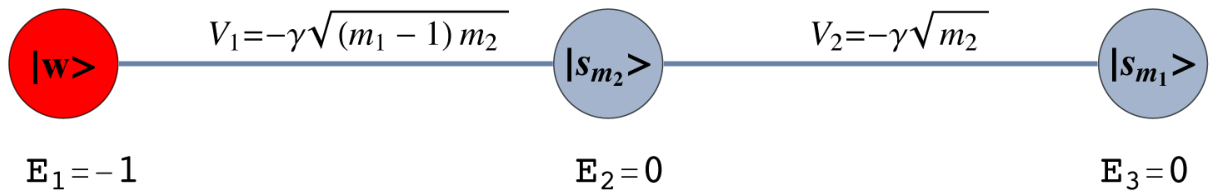}
\caption{}
\label{red_completebigraph_search}
\end{subfigure}
\\~
\vspace{0.5cm}
\\
\begin{subfigure}{0.5\textwidth}
\centering
\includegraphics[scale=0.65,trim= 0cm 0pt 0pt 0pt]{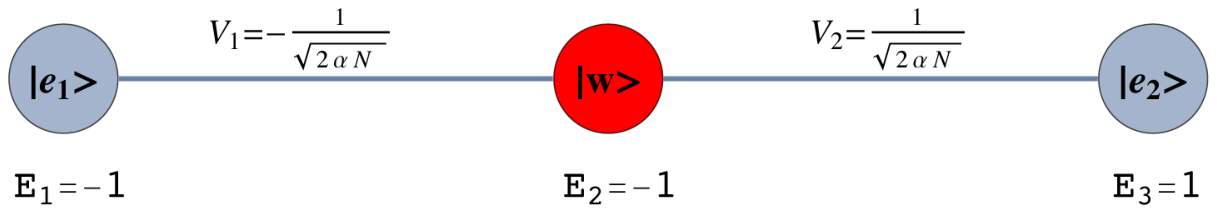}
\caption{}
\label{red_completebigraph_search2}
\end{subfigure}
\caption{ a) Complete bipartite graph $K_{4,3}$ with the solution node $\ket{w}$, represented in red. b) The reduced search Hamiltonian for the complete bipartite graph $K_{m_1,m_2}$ with $m_1+m_2=N$ in the Lanczos basis. $\ket{s_{m_1}}, \ket{s_{m_2}}$ are the equal superposition of the nodes in partition 1 (excluding $\ket{w}$) and 2, respectively. However, the understanding of why the search is optimal in this graph is shown in Fig.~3c. c) The same Hamiltonian as in Fig. 3b, after a basis rotation gives us an idea as to why the algorithm works optimally. The resultant basis is $\ket{e_{1}}=(\ket{s_{m_1}}+\ket{s_{m_2}})/\sqrt{2}$, $\ket{w}$ and $\ket{e_{2}}=(\ket{s_{m_1}}-\ket{s_{m_2}})/\sqrt{2}$. The degeneracy between site energies of $\ket{e_1}$ and $\ket{w}$ facilitates transport between these two nodes while transport between $\ket{w}$ and $\ket{s_{m_2}}$ is inhibited by the  energy gap between them (much larger than the coupling $V_2$). Since there is a considerable overlap between the initial superposition of states $\ket{s}$ and $\ket{e_1}$, there is a large probability amplitude at $\ket{w}$ after a time $T=\mathcal{O}(1/V_1)$.  }
\end{figure}

This result can be perceived as an inherent robustness of the algorithm to imperfections in the form of broken links. One could argue that this robustness has to do with the high connectivity of the structure. However, in the following subsection, we give the example of the star graph, a structure with low connectivity where the algorithm runs optimally that is also robust to broken links. Also, in the context of quantum transport, we show that breaking a link from the complete graph can affect severely the dynamics if one starts with a localized initial state. 
\vspace{.5cm}
\\
\textbf{\textit{Optimal spatial search on complete bipartite graphs}: }
Another example of a highly symmetrical structure, that is in general non-regular, is the complete bipartite graph (CBG). Here, we show that spatial search is optimal for this class of graphs. A complete bipartite graph $G(V_1,V_2,E)$ has two sets of vertices $V_1$ and $V_2$ such that each vertex of $V_1$ is only connected to all vertices of $V_2$ and vice-versa. This set of graphs is also denoted as $K_{m_1,m_2}$, where $m_1=|V_1|$ and $m_2=|V_2|$ and we have $m_1+m_2=N$. This is a non-regular graph, as long as $m_1\neq m_2$. The complete bipartite graph $K_{4,3}$ is shown in Fig. \ref{completebigraph}.
Quantum search was also analysed in these graphs in the formalism of discrete-time scattering quantum walks in \cite{daniel}. However, in that framework, the algorithm does not run optimally if $m_1\gg m_2$. In this case, although each run of the algorithm takes $\mathcal{O}(\sqrt{m_2})$ time, the same must be repeated, on average, $m_1/m_2$ times to find the solution state with high probability. So, if $m_2$ is of $\mathcal{O}(1)$, then $m_1=\mathcal{O}(N)$ and the total running time is linear in $N$.

In our analysis, we show that the CTQW algorithm works in $\mathcal{O}(\sqrt{N})$ time for all possible values of $m_1$ and $m_2$. To analyse the problem, we first assume, without loss of generality, that the solution state $\ket{w}$ belongs to the set of vertices $V_1$ (we shall eliminate this requirement later). The subspace $\mathcal{I}(H_{\text{CBG}},\ket{w})$ is spanned by $\ket{w}$, $\ket{s_{m_1}}=1/\sqrt{m_1-1}\sum_{i\in V_1, i\neq w}\ket{i}$ and $\ket{s_{m_2}}=1/\sqrt{m_2}\sum_{i\in V_2}\ket{i}$. The reduced Hamiltonian can be written as:
\begin{equation}
H_{red}=\begin{bmatrix}
0 & 0& \sqrt{m_2}\\
0 & 0 & \sqrt{(m_1-1) m_2}\\
\sqrt{m_2} & \sqrt{(m_1-1) m_2} & 0
\end{bmatrix}
\end{equation}
Let $\alpha=m_1/N$, $m_2=(1-\alpha) N$ and, $\alpha N \gg 1$. 
Following the procedure in the previous subsection, we 
calculate $H_{\text{search}}=-\ket{w}\bra{w}-\gamma A$, where A is the adjacency matrix of the CBG, and divide it in terms of $\mathcal{O}(1)$ and $\mathcal{O}(1/\sqrt{N})$ with
\begin{align}\label{H0_CBG}
H^{(0)}=&
\begin{bmatrix}
-1 & 0 & 0\\
0 & 0 & -\gamma N\sqrt{\alpha(1-\alpha)}\\
0 & -\gamma N\sqrt{\alpha(1-\alpha)} & 0
\end{bmatrix},\\ \label{H1_CBG}
H^{(1)}=&
\begin{bmatrix}
0 & 0& -\gamma\sqrt{(1-\alpha)N}\\
0 & 0 & 0\\
-\gamma\sqrt{(1-\alpha)N} & 0 & 0
\end{bmatrix}.
\end{align}
This Hamiltonian can be seen as a line with three nodes as shown in Fig.~\ref{red_completebigraph_search}. In order to use perturbation theory we need to diagonalize $H^{(0)}$. The eigenstates of $H^{(0)}$ are $\ket{w}$ with eigenvalue $\lambda_w=-1$, $\ket{e_1}=(\ket{s_{m_1}}+\ket{s_{m_2}})/\sqrt{2}$ with eigenvalue $\lambda_1=-\gamma N\sqrt{\alpha(1-\alpha)}$ and $\ket{e_2}=(\ket{s_{m_1}}-\ket{s_{m_2}})/\sqrt{2}$ with eigenvalue $\lambda_2=\gamma N\sqrt{\alpha(1-\alpha)}$. 

Since $\ket{e_1}$ has the largest overlap with $\ket{s}$, we choose $\gamma=(N\sqrt{\alpha(1-\alpha)})^{-1}$ such that $\ket{e_1}$ and $\ket{w}$ form a degenerate eigenspace of $H^{(0)}$.  The reduced search Hamiltonian in the eigenbasis of $H^{(0)}$, depicted in Fig.~\ref{red_completebigraph_search2}, gives a clearer idea as to why the search is optimal for this graph. The matrix element responsible for the speed of the search is $\bra{e_1}H\ket{w}=1/\sqrt{2\alpha N}.$ Thus, we obtain the running time:
\begin{equation}
T=\pi\sqrt{\dfrac{\alpha N}{2}}.
\end{equation}  
However, unlike previous cases, the success of the algorithm will not be 1 since this will be given by the overlap of the initial state $\ket{s}$ with $\ket{e_1}$. The dynamics will rotate between $\ket{e_1}$ and $\ket{w}$, leaving $\ket{e_2}$ approximately invariant. Thus, the probability of finding the solution by measuring at time $T$ is   
\begin{equation}
P_{\text{suc.}}=|\braket{e_1|s}|^2=\dfrac{1}{2}+\sqrt{\alpha(1-\alpha)}.
\end{equation}
We have $P_{\text{suc.}}=1$ only if $\alpha=1/2$, in which case the graph is regular. It is important to note that one is unaware of which partition contains the solution state. The optimal measurement time depends on whether the solution node is in the partition of $m_1$ nodes or in the partition of $m_2$ nodes. Depending on this, the optimal measurement time would be $T_1=\pi\sqrt{\alpha N/2}$ and $T_2=\pi\sqrt{(1-\alpha) N/2}$ respectively. Thus, the strategy would be to measure interchangeably at time  $T_1$ and then at time $T_2$ until the solution is found. Thus, in such a scenario the expected running time would be $T=T_1+T_2=\pi\sqrt{N}\{1+2\sqrt{\alpha(1-\alpha)}\}^{1/2}$, thereby preserving the quadratic speed up. In fact, this is an upper bound for the expected running time obtained by neglecting the probability of finding the solution even on measurement at the wrong time.

In the extreme case when $m_1=N-1$ and $m_2=1$ i.e., $\alpha=1-1/N$, we obtain a star graph. In this scenario, the optimal $\gamma$ is given by $\gamma\approx1/\sqrt{N}+\mathcal{O}(1/N)$ and the algorithm works with $P_{\text{suc.}}\approx1/2+\mathcal{O}(1/\sqrt{N})$. Thus on average, we have to repeat the algorithm twice to find the solution. We discuss this case in more detail next.
\\~\vspace{.3cm}\\ 
%
\textbf{\textit{Optimal spatial search on the star graph}:}~~
The case of the star graph is particularly interesting because it is a planar structure with node and link connectivity equal to 1. The structures for which quantum search is known to hold optimally are those with typically high connectivity (complete graphs, hypercubes). The quantum search algorithm also works with full quadratic speed-up on lattices of dimension greater than four \citep{Childs_grover} and in two dimensional lattices with a Dirac point in $\mathcal{O}(\sqrt{N}\text{log}N)$ time \cite{tanner_graphene}. We will focus on the case where $m_1=N-1$, $m_2=1$, i.e. the solution is not contained in the central node of the star graph. The case $m_1=1$, $m_2=N-1$, where the solution is in the central node, is trivial because the graph is biased towards the solution and by starting in state $\ket{s}$, we can measure the solution with probability $\approx 1$, in a time $T=\pi/2$, which does not depend on $N$.

So, let the central node be denoted as $\ket{c}$, and assume $\ket{w}\neq\ket{c}$. Since this is a particular case of the CBG, Eqs.~\eqref{H0_CBG} and~\eqref{H1_CBG} with $\alpha=1-1/N$, $\gamma=1/\sqrt{N}$, yield
\begin{align}
H^{(0)}=&
\begin{bmatrix}
-1 & 0 & 0\\
0 & 0 & -1\\
0 & -1 & 0
\end{bmatrix},\\
H^{(1)}=&
\begin{bmatrix}
0 & 0& -1/\sqrt{N}\\
0 & 0 & 0\\
-1/\sqrt{N} & 0 & 0
\end{bmatrix},
\end{align}
in the basis $\{\ket{w},\ket{s_{N-1}},\ket{c}\}$. The eigenstates of $H_0$ are $\ket{w}$, $\ket{e_1}=(\ket{s_{N-1}}+\ket{c})/\sqrt{2}$ and $\ket{e_2}=(\ket{s_{N-1}}-\ket{c})/\sqrt{2}$. In this basis we have
\begin{align}
H_{search}=&
\begin{bmatrix}
-1 & -1/\sqrt{2N} & 1/\sqrt{2N}\\
-1/\sqrt{2N} & -1 & 0 \\
1/\sqrt{2N}& 0 & 1\\
\end{bmatrix}
\end{align}
Using degenerate perturbation theory, we obtain the ground and first excited eigenstates
\begin{equation}
\ket{\lambda_{\pm}}=\frac{\sqrt{2}\ket{w}\pm (\ket{s_{N-1}}+\ket{c})}{2},
\end{equation}
with energies $E_{\pm}=-1\mp 1/\sqrt{2N}$. The running time is given by $T=\pi \sqrt{\frac{N}{2}}$ and probability of success, for the initial state $\ket{s}$ is
\begin{equation}
P_{\text{suc.}}=|\braket{e_1|s}|^2=\dfrac{1}{2}+\mathcal{O}(1/\sqrt{N}).
\end{equation}
It is interesting to note that the algorithm also works, with probability $1/2$, if one starts the quantum walk at the central node $\ket{c}$, since $|\braket{e_1|s}|^2=1/2+\mathcal{O}(1/\sqrt{N})$. This way, one avoids the cost of preparing the initial superposition of states $\ket{s}$ to run the algorithm.
Furthermore, for the star graph, it is easy to analyse the robustness of the algorithm to imperfections in the graph in the form of broken links since the graph obtained after removing $k$ links is still a star graph. We assume $k$ links can be randomly broken so that we possess no knowledge of the links that were broken, nor of the value of $k$. Furthermore, we consider $k$ constant such that $k\ll N$. We assume that the link containing the solution is not removed; this probability is $\mathcal{O}(1/N)$ and therefore negligible for large $N$. In this scenario, the optimal value of $\gamma$ is $\approx 1/\sqrt{N}+\mathcal{O}(k/N)$, so degenerate perturbation theory is still valid and the running time is:
\begin{equation}
T=\pi\sqrt{\dfrac{N}{2}}+\mathcal{O}(k/N).
\end{equation}
with success probability $1/2$.
Thus, quantum search on the star graph is not only optimal but also is fairly robust to defects to the structure in the form of broken links.

\subsection*{Applications to quantum transport}
Let us consider the dynamics of an exciton in a network with $N$ sites, governed by a tight binding Hamiltonian with nearest neighbour couplings, defined as
\begin{equation}
H_{TB}=\sum_{m=1}^{N}{\epsilon_m\ket{m}\bra{m}}+ \sum_{<m,n>}{V_{mn}(\ket{m}\bra{n}+\ket{n}\bra{m})},
\end{equation}
where $\epsilon_m$ is the site energy at site $m$ and $V_{mn}$ is the coupling between site $m$ and $n$. For our analysis, we shall assume that all the site energies are uniform and thus can be set to zero, simply by an overall energy shift. Also, we assume $V_{mn}=V,~ \forall m,n$ and choose our energy units such that $V=1$. Thus, $H_{TB}$ is nothing but the adjacency matrix of a graph with $N$ nodes, whose links connect nearest neighbours of the network (henceforth, we shall use sites and nodes interchangeably). We consider that in one of the nodes of the graph, there is a trap that absorbs the component of the exciton's wave function at this node at a rate $\kappa$, known as the trapping rate. This model is motivated by the study of exciton transport in natural light harvesting systems \cite{mohseniqw, masoud_enaqt, caruso_enaqt, qbiobook}. To model the trapping dynamics, we introduce the trapping Hamiltonian:
\begin{equation}
H_{\text{trap}}=-i\kappa\ket{trap}\bra{trap}.
\end{equation}
This matrix is anti-hermitian and leads to the expected non-unitary dynamics described above. We consider as figure of merit the efficiency of transport \cite{mohseniqw}, defined as
\begin{equation}
\eta=2\kappa\int_{0}^{\infty}dt~\braket{trap|\rho(t)|trap}
\end{equation}
which gives the probability that the exciton is absorbed at the trap integrated over time. The total Hamiltonian describing the dynamics is given by
\begin{equation}
H=A-i\kappa\ket{trap}\bra{trap},
\end{equation} 
where $A$ is the adjacency matrix of the graph. The scenario assumed here is the ideal one, i.e, there is no disorder in the couplings or site energies of the Hamiltonian nor decoherence during the transport. In this regime, in \cite{caruso_enaqt}, the authors calculate the transport efficiency as the overlap of the initial state with the subspace spanned by the eigenstates of the Hamiltonian that have a non-zero overlap with the trap. Earlier in Sec.~\textit{Methods}, it was stated that this subspace is the same as the invariant subspace $\mathcal{I}(H,\ket{w})$, which can be obtained without diagonalizing the Hamiltonian. So the dynamics is such that the component of the initial condition within the space $\mathcal{I}(H,\ket{trap})$ is absorbed by the trap node whereas the component outside this subspace remains in the network (see proof in  \textit{Supplementary Information}). Thus, computing the transport efficiency boils down to finding the overlap of the initial condition with $\mathcal{I}(H,\ket{trap})$.  

In the following subsections, we give examples of how to analytically calculate transport efficiency on various structures given different initial conditions. We also analyse how transport efficiency can be improved by perturbing the symmetry of the complete graph by breaking one link. Finally, we give numerical evidence that breaking a few links in highly symmetric structures leads to the improvement of transport efficiency, if the initial state is localized at one node.

\vspace{.5cm} 
\textbf{\textit{Calculation of transport efficiencies for some graphs with symmetry}:}

\vspace{0.3cm}
\emph{\textbf{Complete graph:}}~~
As in Sec.~\emph{Methods}, we obtain that the reduced subspace for the complete graph with $N$ nodes is given by 
$\{ \ket{trap},\ket{s_{\bar{t}}}\}$ with 
\begin{equation}
\ket{s_{\bar{t}}}=\frac{1}{\sqrt{N-1}}\sum_{i\neq trap}{\ket{i}}
\end{equation}
The reduced Hamiltonian for the transport problem is: 
\begin{equation}
H_{red}=\left(\begin{array}{c c}
-i \kappa & \sqrt{N-1} \\
\sqrt{N-1}& N-2\end{array}\right)
\end{equation}
in the aforementioned basis. If the initial state is localized at a node $\ket{i}\neq \ket{trap}$, the efficiency is given by 
\begin{equation}
\eta=|\braket{i|s_{\bar{t}}}|^2=\frac{1}{N-1}.
\end{equation}
This way, we recover the result obtained in \cite{caruso_enaqt} without the need to solve the equations of motion or requiring to find the eigenstates of the system.

\vspace{.3cm} 
\emph{\textbf{Binary tree:}}~~
Here we consider our graph to be a binary tree with $l$ levels, where the number of nodes is $2^l-1$. The Hamiltonian of the graph is given by:
\begin{equation}
H_{\text{tree}}=\sum_{i=1}^{2^{l-1}-1}(\ket{i}\bra{2i}+ \ket{i}\bra{2i+1} + h.c.)
\end{equation}
We place the trap at the root of the tree, i.e. $\ket{trap}\equiv \ket{1}$. It is well known that a quantum walk on a binary tree can be reduced to the quantum walk on a line \cite{childsglued} where each node represents a column state
\begin{equation}\label{beinglazy}
\ket{\text{col}~j}=\frac{1}{\sqrt{2^{j-1}}}\sum_{a\in \text{column}~ j} \ket{a}
\end{equation}
These column states are readily obtained by applying Lanczos algorithm with the root node $\ket{1}$ (we define $\ket{\text{col}~1}\equiv 1$) as the initial node. If the initial state of the transport problem is a state $\ket{b}$ localized in column $j$, the transport efficiency is:
\begin{equation}
\eta=|\braket{b| \text{col}~j}|^2=\frac{1}{2^{j-1}}.
\end{equation}
Thus, we find that, in such a localized case, the efficiency decreases exponentially with the distance to the trap.

\vspace{.3cm} 
\emph{\textbf{Hypercube:}}~~
Another highly symmetric structure that appears frequently in the literature is the hypercube in the context of quantum computation, quantum transport and state transfer \cite{farhi, kempehype, kay, leo_disaqt}. Here, we consider transport on a hypercube of dimension $d$ with $2^d$ sites. We label the sites of the hypercube by strings of $d$ bits such that each site is connected to another site if they differ by a single bit flip. This way, the Hamiltonian of the graph can be written as:
\begin{equation}
H_{\text{hyp.}}=\sum_{i=1}^d \sigma_{x}^{(i)}
\end{equation}  
where $\sigma_{x}^{(i)}$ is the Pauli matrix $\sigma_x$ acting on the $i$th bit. Using symmetry arguments, it is shown in \cite{farhi, kempehype} that the dynamics in this structure can be reduced to the subspace spanned by the symmetric states with $k$ bits 1 and $n-k$ bits 0:
\begin{equation}\label{dicke}
\ket{D_k^d}={d \choose k}^{-1/2}\sum_{z_1+z_2+...+z_d=k}\ket{z_1}\ket{z_2}...\ket{z_d},
\end{equation}
also known as Dicke states. This is done in the context of the search problem where the solution state is assumed to be at $\ket{w}=\ket{0}^{\otimes d}$. In this picture, the hypercube can be seen as a chain with $d+1$ nodes. Here we also assume, without loss of generality, that our trap state is $\ket{trap}=\ket{0}^{\otimes d}.$ Applying Lanczos algorithm, we also obtain that the invariant subspace, $\mathcal{I}(H,\ket{trap})$ is spanned by the Dicke states in Eq. \eqref{dicke} without using any symmetry arguments. This implies that, if the initial state is localized at a site $\ket{b_j}$, labelled by a bit string with $j$ bits 1, the transport efficiency to the trap is
\begin{equation}
\eta=|\braket{b_j|D_j^d}|^2={d \choose k}^{-1}.
\end{equation}
It is interesting to note that the efficiency is not a monotonic function of the distance from the trap. 

If we consider the initial condition to be a statistical mixture of all sites we obtain the average efficiency:
\begin{equation}
\bar{\eta}=\frac{1}{2^d}\sum_{j=1}^{d+1}{d \choose k}|\braket{b_j|D_j^d}|^2=\frac{d+1}{2^d},
\end{equation}
reproducing analytically the numerical result of \cite{leo_disaqt} for transport on the hypercube of dimension 4 (in the limit of no disorder and no dephasing).

\vspace{.5cm}  
\textit{\textbf{Improving transport efficiency by removing links from highly symmetric graphs}: }

\vspace{.3cm} 
\emph{\textbf{Complete graph with one broken link:}}~~ For a complete graph with $N$ nodes, the transport efficiency is given by $\eta=\frac{1}{N-1}$ provided the transport begins from a localized state. We find that breaking one link from a complete graph increases the transport efficiency. In fact breaking a link that connects the starting node to the trap makes the efficiency of transport go up to 1. Let $\ket{trap}$ denote the trap node, $\ket{i}$ denote the starting node and $\ket{s_{\overline{it}}}$ be the equal superposition of the remaining nodes. The reduced space $\mathcal{I}(H,\ket{trap})$ is spanned by $\{\ket{trap},\ket{i},\ket{s_{\overline{it}}}\}$ and thus, 
\begin{equation}
\eta=|\braket{i|i}|^2=1.
\end{equation}
This is counter-intuitive, as of all the available links, breaking the link that directly connects the starting node to the trap gives the maximum efficiency. Similarly, it can be shown that removing a link between the initial node and another arbitrary node other than the trap gives $\eta=1/2$.

The above phenomenon can be better understood by calculating the dynamics of the resultant graph. In the reduced picture, the trap is coupled to $\ket{s_{\overline{it}}}$ which is in turn coupled to the starting node $\ket{i}$. The reduced Hamiltonian of the graph in the basis $\{\ket{i},\ket{s_{\overline{it}}},\ket{trap}\}$ is,
\begin{equation}
H= \begin{bmatrix}
0 & \sqrt{N-2} & 0\\
\sqrt{N-2} & N-3 & \sqrt{N-2}\\
0 & \sqrt{N-2} & 0
\end{bmatrix}.\\
\end{equation}
Incorporating the anti-Hermitian term of the trap and considering large $N$, for simplicity, results in,
\begin{equation}
H= \begin{bmatrix}
0 & \sqrt{N} & 0\\
\sqrt{N} & N & \sqrt{N}\\
0 & \sqrt{N} & -i\kappa
\end{bmatrix}.\\
\end{equation}
Let the state of the exciton at time $t$, $\ket{\psi(t)}=A_1\ket{i}+A_2\ket{s_{\overline{it}}}+A_3\ket{trap}$. The dynamics of the system is thus,
\begin{align}
\dot{A_1}&=-i\sqrt{N}A_2,~~~\nonumber \\ 
\dot{A_2}&=-i(\sqrt{N}A_1+NA_2+\sqrt{N}A_3), \nonumber \\
\dot{A_3}&=-i\sqrt{N}A_2+i\kappa A_3.
\end{align}
It is important to notice that due to adiabatic elimination, $\dot{A_2}=0$, resulting in an effectively two level system whose dynamics is governed by $A_1$ and $A_3$. The Schr\"{o}dinger equation simplifies to 
\begin{equation}
\begin{bmatrix}
\dot{A_1}\\
\dot{A_3}
\end{bmatrix}=\underbrace{i\begin{bmatrix}
1 & 1\\
1 &1+i\kappa
\end{bmatrix}}_{H^\prime}\begin{bmatrix}
A_1\\
A_3
\end{bmatrix}.
\end{equation}
The eigenstates and eigenvalues of $H^\prime$ are 
\begin{align}
\ket{\lambda_1}&= 1/\sqrt{2}(z\ket{i}+\ket{trap}),~~~~~\lambda_1=1+\bar{z} \nonumber \\
\ket{\lambda_2}&= 1/\sqrt{2}(-\bar{z}\ket{i}+\ket{trap}),~~~\lambda_2=1-\bar{z}.
\end{align}
Here, $z=1/2(-i\kappa-\sqrt{4-\kappa^2})$. It is worth noting that the two eigenstates are not orthogonal and the corresponding eigenvalues have both real and imaginary parts which is due to the fact that $H^\prime$ is not Hermitian. Now, expressing $\ket{trap}$ and $\ket{i}$ in terms of the two eigenstates enables us to calculate the probability of reaching the trap. Thus,
\begin{equation}
p_{trap}(t)=|\braket{trap|U(\tau)|i}|^2=\frac{e^{-\kappa t}}{\Omega^2}\sin^2(\Omega t).
\end{equation}
Here, $U(t)=e^{-iH^\prime t}$ and $\Omega=\frac{1}{2}\sqrt{4-\kappa^2}$. The transport efficiency $\eta$ in this case is 1.

The trapping time or the transfer time, which is the average time required by the exciton to get absorbed by the trap is another relevant measure in quantum transport \cite{masoud_enaqt}. The trapping time is given by,
\begin{equation}
\tau=\frac{2\kappa}{\eta}\int_{0}^{\infty}dt~t~p_{trap}(t).
\end{equation}
In this scenario, $\eta=1$ and,
\begin{align}
\tau&=\frac{2\kappa}{\Omega^2}\int_{0}^{\infty}dt~t~e^{-\kappa t}\sin^2(\Omega t) \nonumber \\
	&=\frac{1}{\kappa}+\frac{\kappa}{2}.
\end{align}
A closer look at $\tau$ shows that for $\kappa$ very small, there is no trapping at all while for a large $\kappa$, one observes freezing of the evolution of the exciton owing to the quantum Zeno effect. The optimal value of the transfer time is obtained for $\kappa=\sqrt{2}$. This is in accordance to \cite{goldilocks, Shabani}, wherein the authors find that the optimal conditions for transport of an exciton in photosynthetic complexes, are when the time scales of hopping and trapping converge.   

\begin{figure}[h]
\includegraphics[scale=0.12, trim= 0cm 0pt 0pt 0pt]{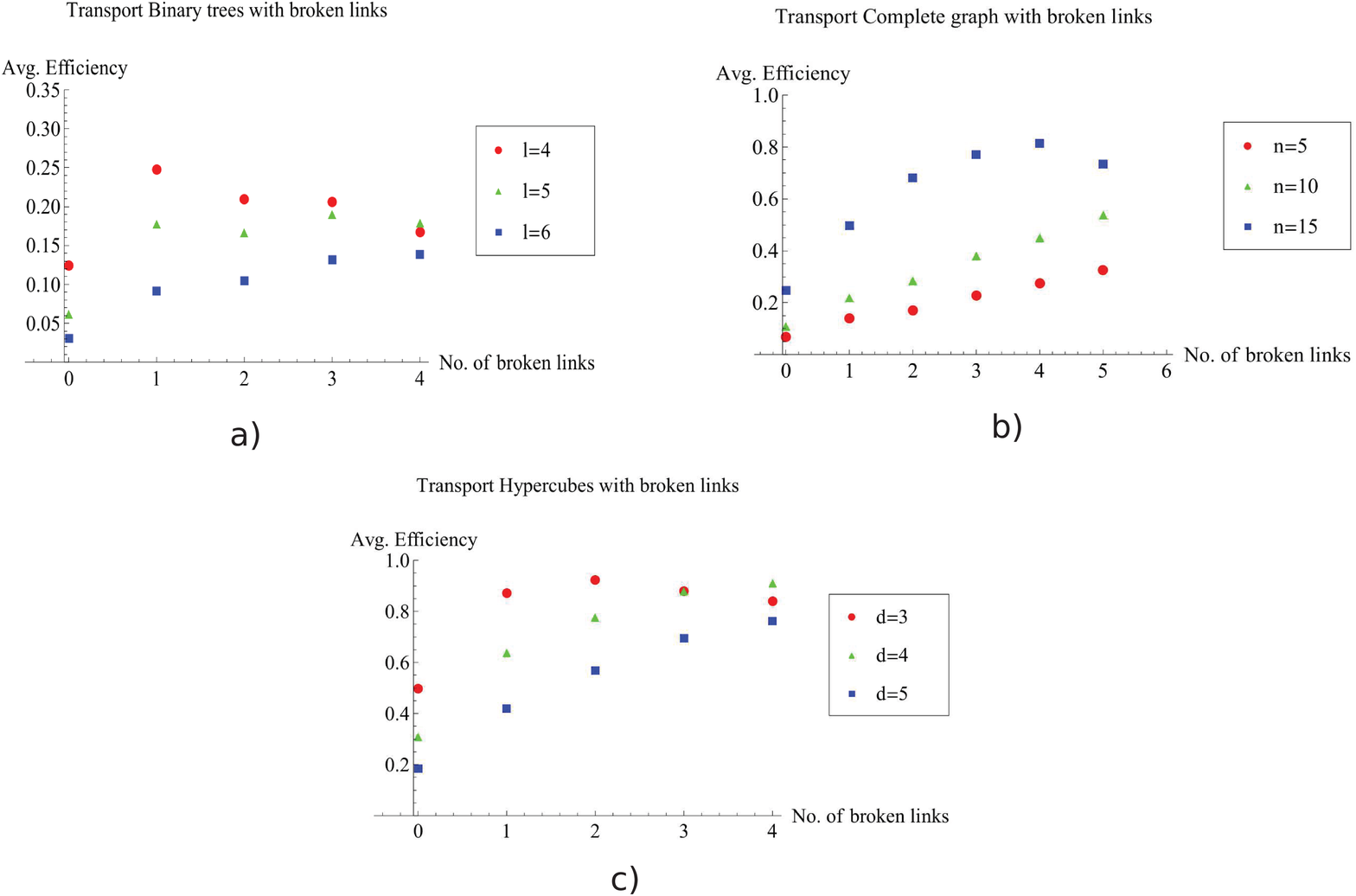}
\caption{Plots of average efficiency versus number of broken links for graphs. a)Binary tree with $l$ levels, b) complete graph with $n$ nodes, c) hypercube with dimension $d$. Clearly, the efficiency obtained by breaking some links is higher than the corresponding efficiency without broken links. This trend continues as long as the number of broken links is not of the order of the total number of links.}
\label{effbl}
\end{figure}

\vspace{.3cm} 
\textit{\textbf{Highly symmetric graphs with broken links:}} Here, we show how the transport efficiency changes as we break links in the graphs mentioned previously in this section. For a graph with $r$ broken links, we calculate the average transport efficiency by projecting the initial state, onto the subspace $\mathcal{I}(H,\ket{trap})$ for each of the possible configurations of the graph with $r$ broken links and average over all of them. The initial state is set as as a statistical mixture of all nodes for the complete graph and the hypercube, while it is a statistical mixture of all leaf nodes in the case of a binary tree. The results are shown in Fig.~\ref{effbl}.

Note that this approach is much faster than diagonalizing the graph Hamiltonian and finding the overlap of the initial state with the eigenstates having non-zero overlap with the trap.

For all these structures, we observe that the average transport efficiency always improves by breaking a few links from the graph. This can be attributed to the fact that by breaking a small number of links, the symmetry is reduced and the dimension of the subspace to which the dynamics is restricted increases. Thus, this can be thought of as increasing the number of possible paths from the starting node to the trap, which previously lied outside this space owing to symmetry. However, we expect that when the number of broken links is comparable to the total number of links in the graph, the size of the reduced space would be very low as the trap gets decoupled from the graph in a large number of configurations, and hence the efficiency is low. This is visible for the case of a binary of three levels and four broken links as shown in Fig.~\ref{effbl} a) with the average transport efficiency being lower than in the case of no broken links. 

\subsection*{Bounds on fidelity of state transfer}
In this section we show that the considerations made thus far for quantum transport can also be applied to the transfer of a qubit state in a network of spins with nearest neighbour interactions. We show that the square root of the efficiencies obtained for transport in various graphs are also upper bounds for the fidelity of the equivalent state transfer problem in the same graph. Thus, all the results obtained for quantum transport can be can also be interpreted in the context of qubit transfer in a network. In particular, we conclude that the fidelity of state transfer can be enhanced by removing links in the network. 

Let us assume we have $N$ spins disposed in the nodes of a graph, where each pair of spins interact if and only if they are connected by an edge. We model the interaction via the $XY$-Hamiltonian with uniform coupling $J$:
\begin{equation}
H_{XY}=\frac{1}{2}J\sum_{(i,j)\in E(G)}\sigma_x^{(i)}\sigma_x^{(j)}+\sigma_y^{(i)}\sigma_y^{(j)},
\end{equation}
where the sum runs over the pairs $(i,j)$ that represent an edge of the graph, and $\sigma_x^{(i)}$ and $\sigma_y^{(i)}$ are Pauli matrices acting on the $i$th spin. The Hamiltonian $H_{XY}$ can be written equivalently as 
\begin{equation}
H_{XY}=J\sum_{(i,j)\in E(G)}\sigma_+^{(i)}\sigma_-^{(j)}+\sigma_-^{(i)}\sigma_+^{(j)}, 
\end{equation}  
using the spin ladder operators $\sigma_+^{(i)}=(\sigma_x^{(i)}+i \sigma_y^{(i)})/2$ and $\sigma_-^{(i)}=(\sigma_x^{(i)}-i \sigma_y^{(i)})/2$. 
This Hamiltonian conserves the number of excitations, i.e. it commutes with the operator $S_z=\sum_{i=1}^N \sigma_z^{(i)}$. If we restrict ourselves to the single excitation subspace of the total Hilbert space, and define the basis 
\begin{equation}
\ket{j}=\ket{0}^{\otimes j-1}\otimes \ket{1}\otimes\ket{0}^{\otimes N-j},
\end{equation}
we can write
\begin{equation}
H_{XY}^{\{\text{1 exc.}\}}=\sum_{(i,j)\in E(G)} \ket{i}\bra{j}+ \ket{j}\bra{i},
\end{equation}
which is the same Hamiltonian defining a continuous time quantum walk used previously. This way, if our initial state is $\ket{i}$, the fidelity of transfer to a node $\ket{w}$ is upper bounded by the overlap of $\ket{i}$ with the subspace $\mathcal{I}(H_{XY},\ket{w})$, since this is an invariant subspace of the Hamiltonian. This way,
\begin{equation}
\mathcal{F}_{\text{max}}=\text{max}_t|\bra{w}U(t)\ket{i}|\leq |\bra{i} P^{\mathcal{I}(H_{XY},\ket{w})}\ket{i}|, 
\end{equation}
where $P^{\mathcal{I}(H_{XY},\ket{w})}$ is the projection operator onto $\mathcal{I}(H_{XY},\ket{w})$. But as $|\bra{i} P^{\mathcal{I}(H_{XY},\ket{w})}\ket{i}|<\sqrt{\eta_i}$ (the efficiency of transport in the graph defined by $H_{XY}$ starting from a localized state $\ket{i}$), the fidelity of transfer to state $\ket{i}$ is bounded by $\sqrt{\eta_i}$. It is important to note that the bound is, in general, not tight. The bound will only be tight when the reduced Hamiltonian ($H_{XY}$ projected onto $\mathcal{I}(H_{XY},\ket{w})$) is the same as the reduced Hamiltonian of a graph where there is perfect state transfer (PST), i.e., a graph where the maximum fidelity of state transfer is one \cite{kay}. Also,  in the cases where the reduced Hamiltonian is the same as the reduced Hamiltonian of a graph where there is pretty good state transfer (PGST),~i.e., the maximum fidelity, $\mathcal{F}_{\text{max}}=|\bra{i} P^{\mathcal{I}(H_{XY},\ket{w})}\ket{i}|-\epsilon$, where $\epsilon$ can be arbitrarily close to 0, the bound is arbitrarily tight \cite{PGST}. This can be fulfilled, for example, if the reduced Hamiltonian is a line having number of nodes $N$ equal to $p-1, 2p-1$, where $p$ is prime, or $2^m-1$ with $m \in \mathbb{N}$. A graph where this is observed is a binary tree with with $l$ levels such that $l$ fulfils these criteria. With this observation, one could think of CTQWs as a way to prepare some multipartite entangled states with high fidelity, which is, in general, a difficult task. A quantum walk on the binary tree starting at the root node (i.e., $\ket{\text{col}~1}=\ket{1}$ ), would evolve, after some time, to a state arbitrarily close to a $W$-state
\begin{equation}
\ket{\text{col}~l}=\frac{1}{\sqrt{2^{l-1}}}\sum_{j\in \text{column}~ l} \ket{j}
\end{equation}
with $\ket{j}$ defined in Eq.~\eqref{beinglazy}. This can be perceived as a way to prepare genuine multipartite entangled states with no time dependent control.

Another way to create such a highly entangled state is by tuning the couplings and site energies of the complete graph as in the spatial search (see Eq.~\ref{Ham_line2d}) such that the dynamics oscillates between a special node, with energy $-1$, and the equal superposition of all other nodes as depicted in Fig.~\ref{red_completegraph_search}. Thus, by starting the quantum walk at this special node, after a time $T= \pi \frac{\sqrt{N}}{2}$, the quantum walk would be in the highly entangled state $\ket{s_{\bar{w}}}$ (see Eq. \ref{swbar}). A physical implementation of a complete graph can be achieved in ions traps where the interaction between the ions can be approximately distance independent \cite{rainer}.

\section*{Discussion}\label{conc}
In this work, we explore the notion of invariant subspaces to simplify the analysis of continuous time quantum walk (CTQW) problems, where the quantity of interest is the probability amplitude  at a particular node of the graph. This way, we obtain new results concerning the spatial search algorithm, quantum transport and state transfer.

First, we present an intuitive picture of the spatial search algorithm by mapping it to a transport problem on a reduced graph whose nodes represent the basis elements of the invariant subspace. Furthermore, we show that the algorithm runs optimally (in $\mathcal{O}(\sqrt{N})$ time) on the complete graph with broken links and on complete bipartite graphs (CBG). These constitute one of the first examples of non-regular graphs where this happens. A particular case of the CBG is the star graph, which is planar, has low connectivity and is robust to imperfections in the form of missing links. Presently, we are considering the robustness of this algorithm to other kinds of defects. During the completion of this article, we came across Refs.~\cite{meyer2, wong}. In the former, it is shown that high connectivity is not a good indicator for optimal spatial search by giving an example of a graph with low connectivity where the algorithm runs optimally, and another graph with high connectivity, where the running time is not optimal. In \cite{wong}, a diagrammatic picture of the spatial search algorithm is presented. 

Furthermore, we present  a simple method to calculate transport efficiency in graphs without having to diagonalize the Hamiltonian. The efficiency is given by the overlap of the initial state with the invariant subspace. Thus, we calculate analytically the transport efficiency in structures such as the complete graph, binary tree and hypercube, given various initial conditions. Moreover, we explore the change in transport efficiency with broken links in these graphs. For the complete graph, breaking a link from the starting node increases the efficiency from $1/(N-1)$ to a constant: $1$, if the link broken was connected to the trap and $1/2$ otherwise. In the former case, we analytically calculate the transfer time which is independent of $N$ and is a function of the trapping rate.

Finally, we show that the square root of the efficiency of transport on a graph from a starting node to a destination (trap) node gives an upper bound on the fidelity of a single qubit transfer between these two nodes. This bound is tight if and only if the reduced Hamiltonian is that of a spin network wherein perfect state transfer takes place.

In summary, dimensionality reduction is an intuitive way to understand the behaviour of CTQWs in graphs with symmetry. Hence, this might lead to the design of new continuous time algorithms, the analysis of the robustness of CTQW algorithms to imperfections, and to novel state transfer and state engineering protocols.

\begin{acknowledgements}
LN, SC and YO thank the support from Funda\c{c}\~{a}o para a Ci\^{e}ncia e a Tecnologia (Portugal), namely through programmes PTDC/POPH and projects PEst-OE/EGE/UI0491/2013, PEst-OE/EEI/LA0008/2013, UID/EEA/50008/2013, IT/QuSim and CRUP-CPU/CQVibes, partially funded by EU FEDER, and from the EU FP7 projects LANDAUER (GA 318287) and PAPETS (GA 323901). MM and HN acknowledge support from Quantum Artificial Intelligence Laboratory at Google. Furthermore, LN and SC acknowledge the support from the DP-PMI and FCT (Portugal) through scholarships SFRH/BD/52241/2013 and SFRH/BD/52246/2013 respectively. 
We would like to thank Mohan Sarovar, Akshat Kumar and Bruno Mera for useful comments, as well as Abolfazal Bayat and Rainer Blatt for useful discussions while visiting the Physics of Information Group in Lisbon. 
\end{acknowledgements}
\section*{Supplementary Material}
\section*{Proof of the equality of $\mathcal{I}(H,\ket{w})$ and $\Lambda(H,\ket{w})$}
In \cite{caruso_enaqt}, the authors calculate the transport efficiency of structures in the no disorder, no dephasing regime, by calculating the overlap of the initial state with the subspace spanned by the eigenstates of the Hamiltonian that have a non-zero overlap with the trap node $\ket{w}$. Let this subspace be denoted by $\Lambda(H,\ket{w})$. Here we prove that this subspace is equal to the space containing the trap node $\ket{w}$ and that is invariant under the unitary evolution, denoted by $\mathcal{I}(H,\ket{w})$. 

Now, $\Lambda(H,\ket{w})=\text{span}(\{\ket{\lambda_1},\dots,\ket{\lambda_m}\})$ where, $\ket{\lambda_k}$ are the minimum number of eigenstates of $H$ with $H\ket{\lambda_k}=\lambda_k\ket{\lambda_k}$, such that $\braket{\lambda_k|w}\neq0$ (there will be $N-m$ eigenstates $\ket{\lambda_k}$ with $\braket{\lambda_k|w}=0$ for $k\in \{m+1,\dots, N\}$ ). Here by \textit{minimum} number of eigenstates, it is meant that in the case of degenerate eigenspaces, more than one eigenstate can have a non-zero overlap with $\ket{w}$. In such a scenario, this ambiguity is resolved by  choosing the eigenvector from this degenerate eigenspace that has the maximum possible overlap with $\ket{w}$ and orthogonalize all the other vectors within this eigenspace with respect to it. This implies that the remaining eigenvectors is the degenerate space would have zero overlap with $\ket{w}$ post orthogonalization. This procedure is explained in \cite{caruso_enaqt} where this subspace is referred to as the non-invariant subspace and its calculation provides a simple way of obtaining the efficiency of transport to a trapping site on the graph (in the absence of dephasing and losses).
\\
Let us first assume that $|\mathcal{I}(H,\ket{w})|=m_1$ and $|\Lambda(H,\ket{w})|=m_2$. It is simple to see that $\mathcal{I}(H,\ket{w})\in\Lambda(H,\ket{w})$ by expressing the state $H^i\ket{w}$ as 
\begin{align}
H^i\ket{w}= &\sum_{k=1}^N\braket{\lambda_k|w}H^i\ket{\lambda_k}\\&=\sum_{k=1}^{m_2}\braket{\lambda_k|w}H^i\ket{\lambda_k} \\&=\sum_{k=1}^{m_2}\braket{\lambda_k|w}{\lambda_k}^i\ket{\lambda_k},
\end{align}
where in the first step we used that $\braket{\lambda_k|w}=0$ for $k\in\{m_2+1,\dots,N\}$.    
Since the states $H^i\ket{w}$ span $\mathcal{I}(H,\ket{w})$ and each of these states can be expressed in terms of elements of $\Lambda(H,\ket{w})$, we conclude that $\mathcal{I}(H,\ket{w})\in\Lambda(H,\ket{w})$ and $m_2\geq m_1$. 

Now, it remains to show that each element of $\Lambda(H,\ket{w})$ can be expressed as
\begin{align}\label{eq:proof1}
\ket{\lambda_j}&=\sum_{i=1}^{m_1}c_{ji}H^{i-1}\ket{w}\\
&=\sum_{i=1}^{m_1}\sum_{k=1}^{m2}c_{ji}\braket{\lambda_k|w}{\lambda_k}^{i-1}\ket{\lambda_k}
\end{align}  
where $c_{ji}$ are coefficients and $j\in\{1,\dots,m_2\}$. For this to happen we obtain the condition
\begin{equation}\label{eq:condition}
\sum_{i=1}^{m_1}c_{ji}\braket{\lambda_k|w}{\lambda_k}^{i-1}=\delta_{jk}.
\end{equation}
Defining the matrix $M_{ik}=\braket{\lambda_k|w}{\lambda_k}^{i-1}$, Eq.~\eqref{eq:condition} is equivalent to the condition 
$\sum_{i=1}^{m_1}c_{ji}M_{ik}=\delta_{jk}$. Thus, $M$ must be an invertible $m\times m$ matrix, so we must have $m_1=m_2=m$. To show that $M$ is always invertible we show that $det(M)\neq0$. For this we define two $m\times m$ matrices, $V_{ij}={\lambda_j}^{i-1}$ and the diagonal matrix $D_{ij}=\delta_{ij}\braket{\lambda_j|w}$ such that $M=D V$. Because $\braket{\lambda_j|w}\neq0$ for $k\in\{1,\dots,m\}$ we conclude that $\text{det}(D)\neq0$. Also, $V$ is of the Vandermonde form so its determinant is given by $\text{det}(V)=\prod_{1\leq i<j\leq m}(\lambda_i-\lambda_j)$. Because all $\ket{\lambda_k}$ belong to different eigenspaces for $k\in\{1,\dots,m\}$, all $\lambda_k$ are different from each other and the Vandermonde determinant is not zero. Thus $\text{det}(M)=\text{det}(D)\text{det}(V)\neq0$ so $M$ is invertible. This completes the proof that $\mathcal{I}(H,\ket{w})=\Lambda(H,\ket{w})$.

Thus, the subspace comprising of the eigenstates of the Hamiltonian having a non-zero overlap with $\ket{w}$ is the same as the one containing $\ket{w}$ and is invariant under the unitary evolution. The advantage of working with the latter subspace is that, one does not need to diagonalize the Hamiltonian to obtain this space.
 
\section*{Optimal search on a complete graph with $k$ broken links such that a link connected to the solution node is broken}

From a complete graph, $k$ links are broken in a manner such that at most one link is broken per node, including the solution state, and thus, $k\leq \frac{N}{2}$. The system Hamiltonian evolves in a four dimensional subspace as we shall show subsequently. We assume that the link connected to the marked state that is broken is represented by $(w,a)$. The remaining $k-1$ broken links are not connected to $w$ and so the set of all these broken links are represented by $E_{\text{broken}}$, whose cardinality is $2k-2$. Also, let $V_{\text{broken}}$ be the set of nodes comprising of these $k-1$ broken links. Now, let
\begin{equation}
\ket{s_{k-1}}=\frac{1}{\sqrt{2k-2}}\sum_{l\in V_{\text{broken}}}\ket{l},
\end{equation}
be the equal superposition of the $2k-2$ nodes corresponding to the $k-1$ broken links that that are not connected to $\ket{w}$. Also let, 
\begin{equation}
\ket{s_{\overline{k-1}}}=\frac{1}{\sqrt{N-2k}}\sum_{\substack{g\notin V_{\text{broken}},\\
g\notin \{\ket{w},\ket{a}\}}}\ket{g},
\end{equation} be the equal superposition of the nodes that have degree $N$, i.e., they do not correspond to any broken link. 

Projecting on to the space $\mathcal{I}(H,\ket{w})$ gives the reduced  Hamiltonian in the basis $\{\ket{w}, \ket{r_{\bar{a}}}, \ket{r_{\bar{a}}^\perp},\ket{a} \}$, where 
\begin{equation}
\ket{r_{\bar{a}}}=\sqrt{\frac{N-2k}{N-2}}\ket{s_{\overline{k-1}}}+\sqrt{\frac{2k-2}{N-2}}\ket{s_{k-1}},
\end{equation} 
and,
 \begin{equation}
 \ket{r_{\bar{a}}^\perp}=\sqrt{\frac{2k-2}{N-2}}\ket{s_{\overline{k-1}}}-\sqrt{\frac{N-2k}{N-2}}\ket{s_{k-1}}.
 \end{equation}
The search Hamiltonian is thus, 
\begin{equation}
H_{\text{search}}= -\gamma
\begin{bmatrix}
\frac{1}{\gamma} & \sqrt{N-2} & 0 & 0\\
\sqrt{N-2} & N+5-\frac{2k+2}{N-2} & \frac{\sqrt{(k-1)(N-2k)}}{N-2} & \sqrt{N-2}\\
0 &  \frac{\sqrt{(k-1)(N-2k)}}{N-2} & -10+\frac{2k+2}{N-2} & 0\\
0 &\sqrt{N-2} & 0 & 0
\end{bmatrix}.
\end{equation}
The initial superposition of states can be approximated to be $\ket{r_{\bar{a}}}$ in the limit of large $N$. Now, let $k=\alpha N$, where, $0\leq\alpha\leq\frac{1}{2}$ and $N$ being large. Thus $H_{\text{search}}$ becomes
\begin{equation}
H_{\text{search}}= -\gamma
\begin{bmatrix}
\frac{1}{\gamma} & \sqrt{N} & 0 & 0\\
\sqrt{N} & N & \sqrt{\alpha(1-2\alpha)} & \sqrt{N}\\
0 & \sqrt{\alpha(1-2\alpha)} & 2\alpha & 0\\
0 & \sqrt{N} & 0 & 0 
\end{bmatrix}.    
\end{equation}
Degenerate perturbation theory enables us to separate $H_{\text{search}}$ into  $H^{(0)}$, $H^{(1)}$ and $H^{(2)}$ of terms of order $\mathcal{O}(1)$, $\mathcal{O}(\frac{1}{\sqrt{N}})$ and $\mathcal{O}(\frac{1}{N})$ respectively. We find the critical value of $\gamma=\frac{1}{N}$ and the eigenvalues of $H^{(0)}+H^{(1)}$ to be $E_\pm=1\pm\frac{1}{\sqrt{N}}$. Thus the running time is again $T=\frac{\pi\sqrt{N}}{2}$, which is the optimal value.

\section*{Transport efficiency for the component of the initial condition within $\mathcal{I}(H,\ket{trap})$}

Here we show that if the initial state $\ket{\psi(0)}\in\mathcal{I}(H,\ket{trap})$, the transport efficiency is one. The efficiency of transporting an exciton from a starting node to the trap is given by
\begin{equation}
\eta={2\kappa}\int_{0}^{\infty}dt~\braket{\psi(t)|trap}\braket{trap|\psi(t)}.
\end{equation}
Now,
\begin{equation}
\frac{d}{dt}(\braket{\psi(t)|\psi(t)})=\braket{\dot{\psi(t)}|\psi(t)}+\braket{\psi(t)|\dot{\psi(t)}}. 
\end{equation}
Using the Schr\"odinger equation to replace $\ket{\dot{\psi(t)}}$,
\begin{equation}
\frac{d}{dt}(\braket{\psi(t)|\psi(t)})=-2\kappa\braket{\psi(t)|trap}\braket{trap|\psi(t)}.
\end{equation}
Thus,
\begin{align}
\eta&=\int_{0}^{\infty}d(\braket{\psi(t)|\psi(t)}) \\  \nonumber
	&=\braket{\psi(0)|\psi(0)}-\braket{\psi(\infty)|\psi(\infty)} \\ \nonumber
	&=1.
\end{align}
This is using the fact that the anti-hermitian term of the Hamiltonian reduces the norm of the state $\ket{\psi(t)}$ at the rate $\kappa$ and for $t\rightarrow \infty$, the component of the wave function within $\mathcal{I}(H,\ket{trap})$ gets absorbed completely and hence $\braket{\psi(\infty)|\psi(\infty)}=0$.

This implies that to calculate the transport efficiency of an exciton starting from an initial state to a trap, it suffices to calculate the overlap of the initial state with $\mathcal{I}(H,\ket{w})$. The component of the exciton outside this subspace will not get absorbed by the trap, but rather will remain in the network. 


\begin{thebibliography}{10}
\expandafter\ifx\csname url\endcsname\relax
  \def\url#1{\texttt{#1}}\fi
\expandafter\ifx\csname urlprefix\endcsname\relax\def\urlprefix{URL }\fi
\providecommand{\bibinfo}[2]{#2}
\providecommand{\eprint}[2][]{\url{#2}}

\bibitem{zagury}
\bibinfo{author}{Aharonov, Y.}, \bibinfo{author}{Davidovich, L.} \&
  \bibinfo{author}{Zagury, N.}
\newblock \bibinfo{title}{Quantum random walks}.
\newblock \emph{\bibinfo{journal}{Physical Review A}}
  \textbf{\bibinfo{volume}{\textbf{48}}}, \bibinfo{pages}{1687}
  (\bibinfo{year}{1993}).

\bibitem{farhidectree}
\bibinfo{author}{Farhi, E.} \& \bibinfo{author}{Gutmann, S.}
\newblock \bibinfo{title}{Quantum computation and decision trees}.
\newblock \emph{\bibinfo{journal}{Physical Review A}}
  \textbf{\bibinfo{volume}{\textbf{58}}}, \bibinfo{pages}{915}
  (\bibinfo{year}{1998}).

\bibitem{kempeqw}
\bibinfo{author}{Kempe, J.}
\newblock \bibinfo{title}{Quantum random walks: an introductory overview}.
\newblock \emph{\bibinfo{journal}{Contemporary Physics}}
  \textbf{\bibinfo{volume}{\textbf{44}}}, \bibinfo{pages}{307}
  (\bibinfo{year}{2003}).

\bibitem{aharonov}
\bibinfo{author}{Aharonov, D.}, \bibinfo{author}{Ambainis, A.},
  \bibinfo{author}{Kempe, J.} \& \bibinfo{author}{Vazirani, U.}
\newblock \bibinfo{title}{Quantum walks on graphs}.
\newblock In \emph{\bibinfo{booktitle}{Proceedings of the thirty-third annual
  ACM symposium on Theory of computing}}, \bibinfo{pages}{50}
  (\bibinfo{organization}{ACM}, \bibinfo{year}{2001}).

\bibitem{kendon}
\bibinfo{author}{Kendon, V.}
\newblock \bibinfo{title}{Quantum walks on general graphs}.
\newblock \emph{\bibinfo{journal}{International Journal of Quantum
  Information}} \textbf{\bibinfo{volume}{\textbf{4}}}, \bibinfo{pages}{791}
  (\bibinfo{year}{2006}).

\bibitem{reitznerreview}
\bibinfo{author}{Reitzner, D.}, \bibinfo{author}{Nagaj, D.} \&
  \bibinfo{author}{Buzek, V.}
\newblock \bibinfo{title}{Quantum walks}.
\newblock \emph{\bibinfo{journal}{Acta Physica Slovaca}}
  \textbf{\bibinfo{volume}{\textbf{61}}}, \bibinfo{pages}{603}
  (\bibinfo{year}{2012}).

\bibitem{venegas}
\bibinfo{author}{Venegas-Andraca, S.~E.}
\newblock \bibinfo{title}{Quantum walks: a comprehensive review}.
\newblock \emph{\bibinfo{journal}{Quantum Information Processing}}
  \textbf{\bibinfo{volume}{11}}, \bibinfo{pages}{1015--1106}
  (\bibinfo{year}{2012}).

\bibitem{yasser2w}
\bibinfo{author}{Omar, Y.}, \bibinfo{author}{Paunkovi{\'c}, N.},
  \bibinfo{author}{Sheridan, L.} \& \bibinfo{author}{Bose, S.}
\newblock \bibinfo{title}{Quantum walk on a line with two entangled particles}.
\newblock \emph{\bibinfo{journal}{Physical Review A}}
  \textbf{\bibinfo{volume}{\textbf{74}}}, \bibinfo{pages}{042304}
  (\bibinfo{year}{2006}).

\bibitem{childsmultiversal}
\bibinfo{author}{Childs, A.~M.}, \bibinfo{author}{Gosset, D.} \&
  \bibinfo{author}{Webb, Z.}
\newblock \bibinfo{title}{Universal computation by multiparticle quantum walk}.
\newblock \emph{\bibinfo{journal}{Science}}
  \textbf{\bibinfo{volume}{\textbf{339}}}, \bibinfo{pages}{791}
  (\bibinfo{year}{2013}).

\bibitem{childsDTvsCT}
\bibinfo{author}{Childs, A.~M.}
\newblock \bibinfo{title}{On the relationship between continuous-and
  discrete-time quantum walk}.
\newblock \emph{\bibinfo{journal}{Communications in Mathematical Physics}}
  \textbf{\bibinfo{volume}{\textbf{294}}}, \bibinfo{pages}{581}
  (\bibinfo{year}{2010}).

\bibitem{Childs_grover}
\bibinfo{author}{Childs, A.~M.} \& \bibinfo{author}{Goldstone, J.}
\newblock \bibinfo{title}{Spatial search by quantum walk}.
\newblock \emph{\bibinfo{journal}{Physical Review A}}
  \textbf{\bibinfo{volume}{\textbf{70}}}, \bibinfo{pages}{022314}
  (\bibinfo{year}{2004}).

\bibitem{childsglued}
\bibinfo{author}{Childs, A.~M.} \emph{et~al.}
\newblock \bibinfo{title}{Exponential algorithmic speedup by a quantum walk}.
\newblock In \emph{\bibinfo{booktitle}{Proceedings of the thirty-fifth annual
  ACM symposium on Theory of computing}}, \bibinfo{pages}{59--68}
  (\bibinfo{organization}{ACM}, \bibinfo{year}{2003}).

\bibitem{childs_universal}
\bibinfo{author}{Childs, A.~M.}
\newblock \bibinfo{title}{Universal computation by quantum walk}.
\newblock \emph{\bibinfo{journal}{Physical Review Letters}}
  \textbf{\bibinfo{volume}{\textbf{102}}}, \bibinfo{pages}{180501}
  (\bibinfo{year}{2009}).

\bibitem{mohseniqw}
\bibinfo{author}{Mohseni, M.}, \bibinfo{author}{Rebentrost, P.},
  \bibinfo{author}{Lloyd, S.} \& \bibinfo{author}{Aspuru-Guzik, A.}
\newblock \bibinfo{title}{Environment-assisted quantum walks in photosynthetic
  energy transfer}.
\newblock \emph{\bibinfo{journal}{The Journal of chemical physics}}
  \textbf{\bibinfo{volume}{129}}, \bibinfo{pages}{174106}
  (\bibinfo{year}{2008}).

\bibitem{masoud_enaqt}
\bibinfo{author}{Rebentrost, P.}, \bibinfo{author}{Mohseni, M.},
  \bibinfo{author}{Kassal, I.}, \bibinfo{author}{Lloyd, S.} \&
  \bibinfo{author}{Aspuru-Guzik, A.}
\newblock \bibinfo{title}{Environment-assisted quantum transport}.
\newblock \emph{\bibinfo{journal}{New Journal of Physics}}
  \textbf{\bibinfo{volume}{\textbf{11}}}, \bibinfo{pages}{033003}
  (\bibinfo{year}{2009}).

\bibitem{pleniodat}
\bibinfo{author}{Plenio, M.~B.} \& \bibinfo{author}{Huelga, S.~F.}
\newblock \bibinfo{title}{Dephasing-assisted transport: quantum networks and
  biomolecules}.
\newblock \emph{\bibinfo{journal}{New Journal of Physics}}
  \textbf{\bibinfo{volume}{\textbf{10}}}, \bibinfo{pages}{113019}
  (\bibinfo{year}{2008}).

\bibitem{caruso_enaqt}
\bibinfo{author}{Caruso, F.}, \bibinfo{author}{Chin, A.~W.},
  \bibinfo{author}{Datta, A.}, \bibinfo{author}{Huelga, S.~F.} \&
  \bibinfo{author}{Plenio, M.~B.}
\newblock \bibinfo{title}{Highly efficient energy excitation transfer in
  light-harvesting complexes: The fundamental role of noise-assisted
  transport}.
\newblock \emph{\bibinfo{journal}{The Journal of Chemical Physics}}
  \textbf{\bibinfo{volume}{\textbf{131}}}, \bibinfo{pages}{105106}
  (\bibinfo{year}{2009}).

\bibitem{sougato_stransfer}
\bibinfo{author}{Bose, S.}
\newblock \bibinfo{title}{Quantum communication through an unmodulated spin
  chain}.
\newblock \emph{\bibinfo{journal}{Physical Review Letters}}
  \textbf{\bibinfo{volume}{\textbf{91}}}, \bibinfo{pages}{207901}
  (\bibinfo{year}{2003}).

\bibitem{kay}
\bibinfo{author}{Kay, A.}
\newblock \bibinfo{title}{Perfect, efficient, state transfer and its
  application as a constructive tool}.
\newblock \emph{\bibinfo{journal}{International Journal of Quantum
  Information}} \textbf{\bibinfo{volume}{\textbf{8}}}, \bibinfo{pages}{641}
  (\bibinfo{year}{2010}).

\bibitem{qbiobook}
\bibinfo{author}{Mohseni, M.}, \bibinfo{author}{Omar, Y.},
  \bibinfo{author}{Engel, G.~S.} \& \bibinfo{author}{Plenio, M.~B.}
\newblock \emph{\bibinfo{title}{Quantum effects in biology}}
  (\bibinfo{publisher}{Cambridge University Press}, \bibinfo{year}{2014}).

\bibitem{severini}
\bibinfo{author}{Bose, S.}, \bibinfo{author}{Casaccino, A.},
  \bibinfo{author}{Mancini, S.} \& \bibinfo{author}{Severini, S.}
\newblock \bibinfo{title}{Communication in xyz all-to all quantum networks with
  a missing link}.
\newblock \emph{\bibinfo{journal}{International Journal of Quantum
  Information}} \textbf{\bibinfo{volume}{\textbf{7}}}, \bibinfo{pages}{713}
  (\bibinfo{year}{2009}).

\bibitem{Meyer_symmetry}
\bibinfo{author}{Janmark, J.}, \bibinfo{author}{Meyer, D.~A.} \&
  \bibinfo{author}{Wong, T.~G.}
\newblock \bibinfo{title}{Global symmetry is unnecessary for fast quantum
  search}.
\newblock \emph{\bibinfo{journal}{Physical Review Letters}}
  \textbf{\bibinfo{volume}{\textbf{112}}}, \bibinfo{pages}{210502}
  (\bibinfo{year}{2014}).

\bibitem{Farhi_analog_grover}
\bibinfo{author}{Farhi, E.} \& \bibinfo{author}{Gutmann, S.}
\newblock \bibinfo{title}{Analog analogue of a digital quantum computation}.
\newblock \emph{\bibinfo{journal}{Physical Review A}}
  \textbf{\bibinfo{volume}{\textbf{57}}}, \bibinfo{pages}{2403}
  (\bibinfo{year}{1998}).

\bibitem{lanczos}
\bibinfo{author}{Lanczos, C.}
\newblock \bibinfo{title}{An iteration method for the solution of the
  eigenvalue problem of linear differential and integral operators}.
\newblock \emph{\bibinfo{journal}{J. Res. Nat'l Bur. Std.}}
  \textbf{\bibinfo{volume}{\textbf{45}}}, \bibinfo{pages}{225}
  (\bibinfo{year}{1950}).

\bibitem{tanner_graphene}
\bibinfo{author}{Foulger, I.}, \bibinfo{author}{Gnutzmann, S.} \&
  \bibinfo{author}{Tanner, G.}
\newblock \bibinfo{title}{Quantum search on graphene lattices}.
\newblock \emph{\bibinfo{journal}{Physical Review Letters}}
  \textbf{\bibinfo{volume}{112}}, \bibinfo{pages}{070504}
  (\bibinfo{year}{2014}).

\bibitem{childs_2d_search}
\bibinfo{author}{Childs, A.~M.} \& \bibinfo{author}{Ge, Y.}
\newblock \bibinfo{title}{Spatial search by continuous-time quantum walks on
  crystal lattices}.
\newblock \emph{\bibinfo{journal}{Physical Review A}}
  \textbf{\bibinfo{volume}{89}}, \bibinfo{pages}{052337}
  (\bibinfo{year}{2014}).

\bibitem{aaronson}
\bibinfo{author}{Aaronson, S.} \& \bibinfo{author}{Ambainis, A.}
\newblock \bibinfo{title}{Quantum search of spatial regions}.
\newblock In \emph{\bibinfo{booktitle}{Proceedings of the 44th Annual IEEE
  Symposium on Foundations of Computer Science}}, \bibinfo{pages}{200}
  (\bibinfo{publisher}{IEEE Computer Society}, \bibinfo{address}{Washington,
  DC, USA}, \bibinfo{year}{2003}).

\bibitem{geometrical}
\bibinfo{author}{Mohseni, M.}, \bibinfo{author}{Shabani, A.},
  \bibinfo{author}{Lloyd, S.}, \bibinfo{author}{Omar, Y.} \&
  \bibinfo{author}{Rabitz, H.}
\newblock \bibinfo{title}{Geometrical effects on energy transfer in disordered
  open quantum systems}.
\newblock \emph{\bibinfo{journal}{The Journal of chemical physics}}
  \textbf{\bibinfo{volume}{\textbf{138}}}, \bibinfo{pages}{204309}
  (\bibinfo{year}{2013}).

\bibitem{zafar}
\bibinfo{author}{Jafarizadeh, M.}, \bibinfo{author}{Sufiani, R.},
  \bibinfo{author}{Salimi, S.} \& \bibinfo{author}{Jafarizadeh, S.}
\newblock \bibinfo{title}{Investigation of continuous-time quantum walk by
  using krylov subspace-lanczos algorithm}.
\newblock \emph{\bibinfo{journal}{The European Physical Journal B-Condensed
  Matter and Complex Systems}} \textbf{\bibinfo{volume}{59}},
  \bibinfo{pages}{199--216} (\bibinfo{year}{2007}).

\bibitem{brun}
\bibinfo{author}{Krovi, H.} \& \bibinfo{author}{Brun, T.}
\newblock \bibinfo{title}{Quantum walks on quotient graphs}.
\newblock \emph{\bibinfo{journal}{Physical Review A}}
  \textbf{\bibinfo{volume}{\textbf{75}}}, \bibinfo{pages}{062332}
  (\bibinfo{year}{2007}).

\bibitem{daniel}
\bibinfo{author}{Reitzner, D.}, \bibinfo{author}{Hillery, M.},
  \bibinfo{author}{Feldman, E.} \& \bibinfo{author}{Bu$\check{z}$ek, V.}
\newblock \bibinfo{title}{Quantum searches on highly symmetric graphs}.
\newblock \emph{\bibinfo{journal}{Physical Review A}}
  \textbf{\bibinfo{volume}{79}}, \bibinfo{pages}{012323}
  (\bibinfo{year}{2009}).

\bibitem{sufiani}
\bibinfo{author}{Sufiani, R.} \& \bibinfo{author}{Bahari, N.}
\newblock \bibinfo{title}{Quantum search in structured database using local
  adiabatic evolution and spectral methods}.
\newblock \emph{\bibinfo{journal}{Quantum Information Processing}}
  \textbf{\bibinfo{volume}{12}}, \bibinfo{pages}{2813--2831}
  (\bibinfo{year}{2013}).

\bibitem{gong}
\bibinfo{author}{Gong, Z.-X.}, \bibinfo{author}{Foss-Feig, M.},
  \bibinfo{author}{Michalakis, S.} \& \bibinfo{author}{Gorshkov, A.~V.}
\newblock \bibinfo{title}{Persistence of locality in systems with power-law
  interactions}.
\newblock \emph{\bibinfo{journal}{Physical Review Letters}}
  \textbf{\bibinfo{volume}{113}}, \bibinfo{pages}{030602}
  (\bibinfo{year}{2014}).

\bibitem{mohan}
\bibinfo{author}{Kumar, A.} \& \bibinfo{author}{Sarovar, M.}
\newblock \bibinfo{title}{On model reduction for quantum dynamics: symmetries
  and invariant subspaces}.
\newblock \emph{\bibinfo{journal}{Journal of Physics A: Mathematical and
  Theoretical}} \textbf{\bibinfo{volume}{\textbf{48}}}, \bibinfo{pages}{015301}
  (\bibinfo{year}{2015}).

\bibitem{farhi}
\bibinfo{author}{Farhi, E.}, \bibinfo{author}{Goldstone, J.},
  \bibinfo{author}{Gutmann, S.} \& \bibinfo{author}{Sipser, M.}
\newblock \bibinfo{title}{Quantum computation by adiabatic evolution}.
\newblock \emph{\bibinfo{journal}{quant-ph/0001106}}  (\bibinfo{year}{2000}).

\bibitem{kempehype}
\bibinfo{author}{Kempe, J.}
\newblock \bibinfo{title}{Discrete Quantum Walks Hit Exponentially Faster}.
\newblock \emph{\bibinfo{journal}{Probability Theory and Related Fields}} \textbf{\bibinfo{volume}{\textbf{133}}}, \bibinfo{pages}{215--235}
  (\bibinfo{year}{2005}).

\bibitem{leo_disaqt}
\bibinfo{author}{Novo, L.}, \bibinfo{author}{Mohseni, M.} \&
  \bibinfo{author}{Omar, Y.}
\newblock \bibinfo{title}{Disorder-assisted quantum transport in suboptimal
  decoherence regimes}.
\newblock \emph{\bibinfo{journal}{arXiv:1312.6989}}  (\bibinfo{year}{2013}).

\bibitem{goldilocks}
\bibinfo{author}{Mohseni, M.}, \bibinfo{author}{Shabani, A.},
  \bibinfo{author}{Lloyd, S.} \& \bibinfo{author}{Rabitz, H.}
\newblock \bibinfo{title}{Energy-scales convergence for optimal and robust
  quantum transport in photosynthetic complexes}.
\newblock \emph{\bibinfo{journal}{The Journal of Chemical Physics}}
  \textbf{\bibinfo{volume}{\textbf{140}}}, \bibinfo{pages}{035102}
  (\bibinfo{year}{2014}).

\bibitem{Shabani}
\bibinfo{author}{Shabani, A.}, \bibinfo{author}{Mohseni, M.},
  \bibinfo{author}{Rabitz, H.} \& \bibinfo{author}{Lloyd, S.}
\newblock \bibinfo{title}{Numerical evidence for robustness of
  environment-assisted quantum transport}.
\newblock \emph{\bibinfo{journal}{Physical Review E}}
  \textbf{\bibinfo{volume}{\textbf{89}}}, \bibinfo{pages}{042706}
  (\bibinfo{year}{2014}).

\bibitem{PGST}
\bibinfo{author}{Godsil, C.}, \bibinfo{author}{Kirkland, S.},
  \bibinfo{author}{Severini, S.} \& \bibinfo{author}{Smith, J.}
\newblock \bibinfo{title}{Number-theoretic nature of communication in quantum
  spin systems}.
\newblock \emph{\bibinfo{journal}{Physical Review Letters}}
  \textbf{\bibinfo{volume}{\textbf{109}}}, \bibinfo{pages}{050502}
  (\bibinfo{year}{2012}).

\bibitem{rainer}
\bibinfo{author}{Jurcevic, P.} \emph{et~al.}
\newblock \bibinfo{title}{Quasiparticle engineering and entanglement
  propagation in a quantum many-body system}.
\newblock \emph{\bibinfo{journal}{Nature}}
  \textbf{\bibinfo{volume}{\textbf{511}}}, \bibinfo{pages}{202}
  (\bibinfo{year}{2014}).

\bibitem{meyer2}
\bibinfo{author}{Meyer, D.~A.} \& \bibinfo{author}{Wong, T.~G.}
\newblock \bibinfo{title}{Connectivity is a poor indicator of fast quantum
  search}.
\newblock \emph{\bibinfo{journal}{Physical Review Letters}}
  \textbf{\bibinfo{volume}{114}}, \bibinfo{pages}{110503}
  (\bibinfo{year}{2015}).

\bibitem{wong}
\bibinfo{author}{Wong, T.}
\newblock \bibinfo{title}{Diagrammatic approach to quantum search}.
\newblock \emph{\bibinfo{journal}{Quantum Information Processing}} 
 \textbf{\bibinfo{volume}{14}},
  \bibinfo{pages}{1767--1775} (\bibinfo{year}{2015}).

\end{thebibliography}
\end{document}